\documentclass[12pt]{svjour2}

\smartqed
\usepackage{graphics,graphicx,subfigure,adjustbox,amssymb,xcolor}
\usepackage{amsmath}

\journalname{J Stat Phys}

\begin{document}

\title{Counterions between walls with surface charge modulations: Exact Poisson-Boltzmann solutions}

\titlerunning{Counterions between walls with surface charge modulations}

\author{Ladislav \v{S}amaj}

\institute{Institute of Physics, Slovak Academy of Sciences, 
D\'ubravsk\'a cesta 9, SK-84511 Bratislava, Slovakia \\
\email{Ladislav.Samaj@savba.sk}}

\date{Received:  / Accepted: }

\maketitle

\begin{abstract}
We study the thermal equilibrium of a classical system of identical point
charges moving between two walls with parallel surfaces at a distance $d$,
charged symmetrically with position-dependent surface charge densities.
Specifically, we study the effect of surface charge modulation on
the effective force (pressure) between the walls in the Poisson-Boltzmann
limit.
Based on Monte Carlo simulations at high temperatures, it is predicted that
surface charge modulation reduces the pressure between the walls compared
to uniformly charged wall surfaces with the same average surface
charge density.
We restrict ourselves to surface charge modulations in only one direction
and, using a general solution of the two-dimensional Liouville equation,
we construct exact solutions for the electrostatic potential.
The surface charge density on the walls is generated
inversely from this potential, which means that our exact results apply
to a limited set of models with a specific variation of the surface charge
density with distance $d$.
Explicit analytical results show that surface charge modulation can both
increase (small distances $d$) and decrease (large $d$) the pressure
between the walls. 


\end{abstract}

\renewcommand{\theequation}{1.\arabic{equation}}
\setcounter{equation}{0}

\section{Introduction} \label{intro}
As a result of solvation and screening, large colloids immersed in
a polar solvent such as water release mobile particles from their surfaces
which are ``counterions'' with respect to the surface charge density
created on the colloid surface.
Although mobile ions in Coulomb fluids are generically of both signs,
one can approach experimentally the limit of deionized (or salt-free)
suspensions \cite{Raspaud00,Palberg04,Brunner04}.
Counterions in the vicinity of a charged colloid represent the simplest
example of an electric double layer (EDL) \cite{Attard96,Levin02,Messina09}.
To simplify the theoretical description of the complex Coulomb problem, 
the curved surface of large colloids is usually replaced by a planar one,
the modulated shape of the fixed charge density on the colloid surface
is replaced by a uniform one, and the dielectric jump between the interior
of the colloid and the solvent is ignored.
In general, two geometries are studied: a single EDL and two parallel EDLs. 
The study of the effective interaction between two similarly charged EDLs,
mediated by counterions, is of particular experimental and theoretical interest
\cite{Hansen00} due to the counterintuitive phenomenon of effective
like-charge attraction observed in computer simulations
\cite{Gulbrand84,Kjellander84,Gronbech97} and experiments
\cite{Khan85,Kjellander88,Bloomfield91,Kekicheff93,Dubois98}.
The relative simplicity of the studied models with uniform surface charge
densities allows the development of a systematic treatment of both opposite
weak-coupling \cite{Attard88,Podgornik90,Netz00} and strong-coupling
\cite{Moreira01,Netz01,Boroudjerdi05,Kanduc07,Samaj11,Samaj16} limits.

The first theoretical approaches to EDLs with heterogeneous surface charge
densities were based on liquid-state approximations
\cite{Kjellander88,Chan80,Gonzalez01}.
In the weak-coupling regime, analytic perturbation studies in combination
with Monte Carlo simulations \cite{Lukatsky02a,Henle04} suggest
an increase in the average counterion density near an inhomogeneously
charged surface compared to a surface charged uniformly with the same
average charge density.
In the case of two parallel walls with modulated surface charge densities
the increase in the counterion density at the surfaces is accompanied
by a decrease in the counterion density in the midplane, which
is argued to imply a decrease in the pressure between the walls
\cite{Lukatsky02b,Khan05}.
This phenomenon was indeed observed in a 2D exactly solvable version of
parallel EDLs with modulated surface charge density at the free-fermion
coupling $\Gamma=2$ \cite{Samaj22}.

The contact value theorem for EDLs with uniform surface charge densities
\cite{Henderson78,Henderson79,Blum81,Wennerstrom82} has been generalized
to EDLs with modulated surface charge densities in Ref. \cite{Samaj25}.
Unlike the uniform case, the pressure formula includes not only
the particle density at the wall surface, but also the particle density
profile between the walls; therefore, its application is limited.

In this paper, we focus on the weak coupling regime described by
the Poisson-Boltzmann (PB) mean field theory, initiated
long ago by Gouy \cite{Gouy10} and Chapman \cite{Chapman13}.
Several exact solutions have been found for the counterion density profile
around a single planar EDL with specific forms of surface charge
modulations along only one direction \cite{Samaj19}.
Explicit results for the planar geometry confirm the increase in
the density of counterions on the wall due to the modulation of the surface
charge density.
A similar result can be expected in the case of two symmetric parallel EDLs,
leading to a decrease in the counterion density in the midplane between
the walls.
As we show in this work, although the PB pressure is determined by
the particle density in the midplane between the walls, there is
an additional (positive) contribution to the pressure due to the variation
of the electrostatic potential along the midplane.
This positive contribution can eliminate or even overcome the negative
contribution caused by the decrease in particle density in the midplane
between the walls.

This work extends previous exact solutions for a single EDL with
modulated surface charge density \cite{Samaj19} to a pair of parallel
EDLs at distance $d$, symmetrically charged with the same surface
charge densities modulated in only one direction.
The method is based on the general formula for exact solutions of
the 2D Liouville equation given in Ref. \cite{Crowdy97}.
The task is nontrivial, since it is necessary to choose only physically
acceptable ``trial'' functions that, when substituted into the general formula
for the electrostatic potential, do not produce singularities
(i.e., additional point charges) at any point in the space accessible
to mobile counterions.
Another requirement is that the inhomogeneous surface charge densities
be the same for each of the two wall surfaces.
Since the surface charge density is generated inversely
from the exact electrostatic potential, changing the distance between
the walls $d$ leads to a change in the surface charge density profile.
This means that our exact results apply to a limited set of models with
a specific variation of the surface charge density with distance $d$,
and not to a general model with an arbitrary fixed charge distribution.
The main result of the analysis of our exactly solvable
family is that surface charge modulation increases/decreases
the pressure for small/large wall distances.
This result corrects previous mean-field ideas about the effect of
surface charge modulation on the effective wall interaction.

The paper is organized as follows.
The general formalism of the mean-field PB approach for two parallel EDLs
charged symmetrically with modulated surface charge density
is reviewed in section \ref{meanfield}.
The standard case of uniform surface charge density
is briefly summarized in section \ref{uniform}.
The systematic expansion of the electrostatic potential and counterion
density when the variation of the surface charge density around its mean
value is weak is elaborated in section \ref{infinite}.
Section \ref{exact} deals with the exact solution for the geometry of
two parallel walls.
Section \ref{general} recapitulates the general Crowdy formula \cite{Crowdy97},
which covers all possible solutions of the 2D Liouville equation.
A particular solution of the 2D Liouville equation physically suitable
for the electrostatic potential is discussed in section \ref{particular}.
Section \ref{conditions} summarizes the necessary conditions for
the regularity of the exact solution of the potential.
Section \ref{analysis} discusses the physical models that correspond
to the specific exact solutions of the 2D Liouville equation and provides
an analysis of the obtained exact results.
Emphasis is put on the specific characteristics of the modulated surface
charge densities generated from the exact electrostatic potential and
the effect of the wall spacing and surface charge modulation on the pressure.
Section \ref{conclusion} is a short summary with concluding remarks. 

\renewcommand{\theequation}{2.\arabic{equation}} 
\setcounter{equation}{0}

\section{PB approach to two-wall geometry} \label{meanfield} 

\subsection{General formalism for modulated surface charge density}
In a three-dimensional Euclidean space of points with Cartesian
coordinates ${\bf r}=(x,y,z)$, let there exist a pair of parallel walls
surfaces perpendicular to the $x$ axis at a distance $d$, the left one
being located at $x=-d/2$ and the right one at $x=d/2$.
The longitudinal coordinates $y\in [-L_y/2,L_y/2]$ and $z\in [-L_z/2,L_z/2]$
are unbounded, $L_y,L_z\to\infty$.
Each of the walls carries the same surface charge density $\sigma(y,z) e$,
where $e$ denotes a unit charge.
To obtain exact results, we restrict ourselves to surface charge
densities that vary only in one direction, say $\sigma(y,z) \equiv \sigma(y)$.
Then all quantities considered in this work will depend on coordinates
$x$ and $y$, but not on $z$, i.e. the problem is effectively
two-dimensional (2D).
In the original formulation of the problem, when particles are
released from the colloidal surfaces directly into the solvent,
it holds that $\sigma(y)\ge 0$.
If some particles can return to some locations on the colloidal surfaces,
then this restriction is absent and $\sigma(y)$ can be negative in
certain intervals of the $y$ axis.

\begin{figure}[]
\begin{center}
\includegraphics[clip,width=0.9\textwidth]{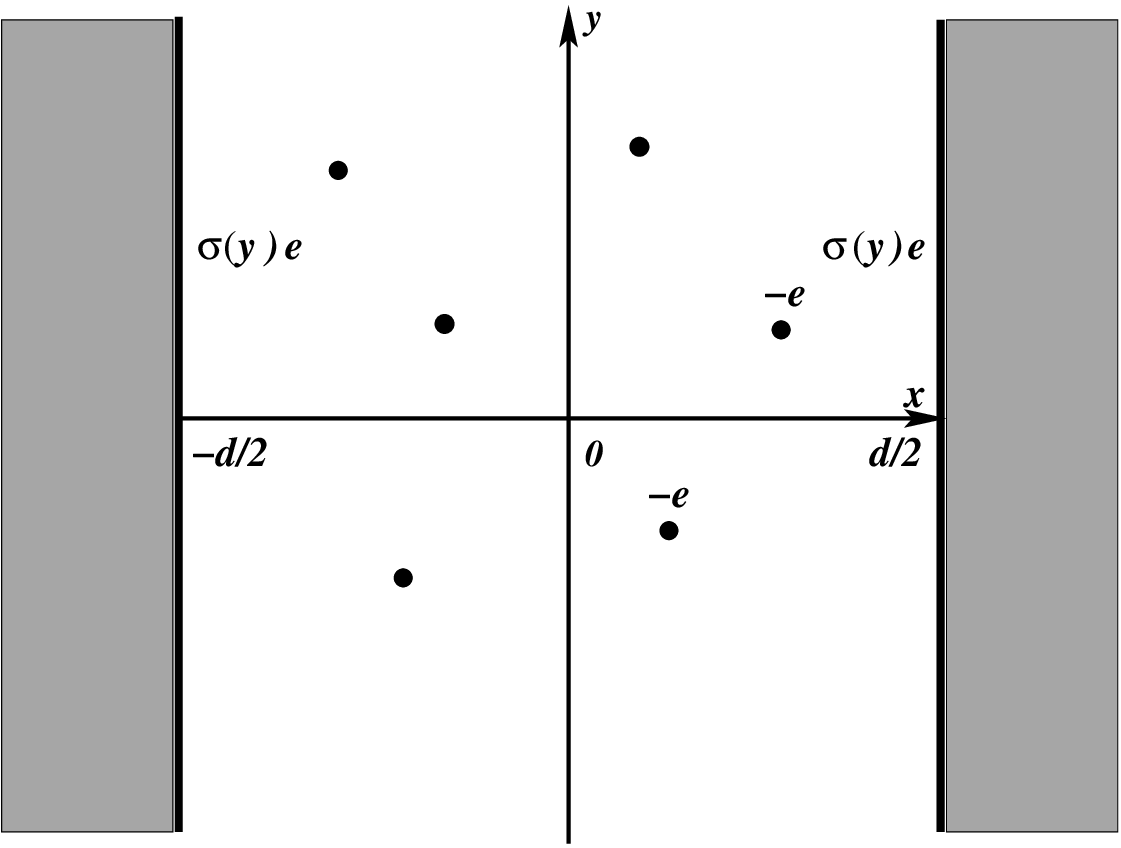}
\caption{Geometry of two parallel walls with surfaces at a distance $d$,
one at $x=-d/2$ and the other at $x=d/2$.
The symmetric surface charge density on the walls $\sigma(y) e$ depends on
the coordinate $y$ perpendicular to $x$.
Point particles with charge $-e$ represented by hard disks
move between the walls.}
\label{fig1}
\end{center}
\end{figure}

Identical point particles of charge $-e$, called ``counterions'',
move inside the region $-d/2<x<d/2$, see Fig. \ref{fig1}.
The particles are immersed in a medium with dielectric constant
$\varepsilon$ and interact in pairs via the Coulomb potential
$1/(\varepsilon r)$ (in Gauss units).
The dielectric constant of the walls is assumed for simplicity to be
the same as the dielectric constant of the medium, i.e. there are
no electrostatic image charges.  
The system of particles is in thermal equilibrium at the temperature $T$,
or the inverse temperature $\beta=1/(k_{\rm B}T)$, where $k_{\rm B}$ is the
Boltzmann constant.

Let $n(x,y)$ be the mean particle density (per unit surface of either
of the walls) and $\rho(x,y) = - e n(x,y)$ be the corresponding particle
charge density at point ${\bf r}$.
The condition for overall electroneutrality is
\begin{equation} \label{neutrality}
2 \int_{-L_y/2}^{L_y/2} {\rm d} y\, \sigma(y) e +
\int_{-d/2}^{d/2} {\rm d} x\, \int_{-L_y/2}^{L_y/2} {\rm d} y\, \rho(x,y) = 0 .
\end{equation}
The mean electrostatic potential $\psi(x,y)$ satisfies Poisson's equation
\cite{Jackson98}
\begin{equation} \label{Poisson}
\frac{\partial^2 \psi}{\partial x^2}  + \frac{\partial^2 \psi}{\partial y^2}
= - \frac{4\pi}{\varepsilon} \rho .
\end{equation}
The normal derivative of $\psi$ on the wall surface is related to the surface
charge density as \cite{Jackson98}
\begin{equation} \label{bc}
\frac{\partial\psi(x,y)}{\partial x}\Big\vert_{x=\pm d/2} = 
\pm \frac{4\pi \sigma(y) e}{\varepsilon} . 
\end{equation}
These boundary conditions are consistent with the neutrality requirement
(\ref{neutrality}), as one can show by integrating Poisson's equation
(\ref{Poisson}) over the entire space between the walls.

There are two relevant lengths.
The distance between two unit charges $e$ immersed in a medium with dielectric
constant $\varepsilon$ at which they interact with thermal energy
$k_{\rm B}T$ is called the Bjerrum length
\begin{equation}
\ell_{\rm B} \equiv \frac{\beta e^2}{\varepsilon} .
\end{equation}
The distance from a wall charged with uniform surface charge density
$\sigma e$ at which the potential energy of a unit charge $e$ equals the
thermal energy $k_{\rm B}T$ is the Gouy-Chapman length \cite{Gouy10,Chapman13}
\begin{equation}
\mu = \frac{1}{2\pi\ell_{\rm B}\sigma} .
\end{equation}
The coupling constant $\Xi$ is the ratio of these two lengths:
\begin{equation}
\Xi = \frac{\ell_{\rm B}}{\mu} .
\end{equation}  
The weak coupling regime $\Xi\ll 1$ is adequately described by the mean-field
PB theory \cite{Gouy10,Chapman13}.
Within the PB theory, the particle density at a point is proportional to
the Boltzmann weight of the mean electrostatic potential (multiplied by
the particle charge $-e$) at that point:
\begin{equation}
n({\bf r}) = f_0 {\rm e}^{\beta e \psi({\bf r})} ,
\end{equation}
where $f_0$ is the normalization constant.
In terms of the reduced potential $\phi = \beta e \psi$, this relation reads as
\begin{equation} \label{density}
n({\bf r}) = f_0 {\rm e}^{\phi({\bf r})} .
\end{equation}

Let us introduce dimensionless coordinates
\begin{equation} \label{coor}
\widetilde{x}= K x , \quad \widetilde{y}= K y ,
\end{equation}
where
\begin{equation} \label{K}
K = \sqrt{2\pi\ell_{\rm B}f_0} 
\end{equation}
is a free normalization parameter for now.
Poisson's equation (\ref{Poisson}) can be rewritten for
the reduced potential as the standard 2D Liouville equation
\begin{equation} \label{Poissonprime}
\frac{\partial^2 \phi}{\partial \widetilde{x}^2}  
+ \frac{\partial^2 \phi}{\partial \widetilde{y}^2}
= 2 {\rm e}^{\phi} 
\end{equation}
and the boundary conditions (\ref{bc}) are expressed as
\begin{equation} \label{bcprime}
\frac{\partial\phi(\widetilde{x},\widetilde{y})}{
\partial \widetilde{x}}\Big\vert_{\widetilde{x}=\pm \widetilde{d}/2} = 
\pm \frac{4\pi \ell_{\rm B}\sigma(\widetilde{y})}{K} . 
\end{equation}
The particle density (\ref{density}) can be expressed using
the dimensionless potential as follows
\begin{equation} \label{densityprime}
n(\widetilde{\bf r}) = \frac{K^2}{2\pi\ell_{\rm B}}
{\rm e}^{\phi(\widetilde{\bf r})} .
\end{equation}

Since for symmetrically charged walls the reduced potential exhibits
the reflection symmetry,
\begin{equation} \label{reflection}
\phi(\widetilde{x},\widetilde{y}) =
\phi(-\widetilde{x},\widetilde{y}) ,
\end{equation}  
the two symmetric boundary conditions (\ref{bcprime}) can be replaced by
the equivalent ones
\begin{equation} \label{bcprimeprime}
\frac{\partial\phi(\widetilde{x},\widetilde{y})}{
\partial \widetilde{x}}\Big\vert_{\widetilde{x}=0} = 0 , \qquad
\frac{\partial\phi(\widetilde{x},\widetilde{y})}{
\partial \widetilde{x}}\Big\vert_{\widetilde{x}=\widetilde{d}/2} = 
\frac{4\pi \ell_{\rm B}\sigma(\widetilde{y})}{K} . 
\end{equation}
These boundary conditions formally correspond to the configuration of one wall
without surface charge density localized at $\widetilde{x}=0$ and
the other wall at $\widetilde{x}=\widetilde{d}/2$ with
inhomogeneous surface charge density $\sigma(\widetilde{y})$.
The electroneutrality condition (\ref{neutrality}) has the form
\begin{equation} \label{neutralityprime}
K \int_{-\widetilde{L}_y/2}^{\widetilde{L}_y/2} {\rm d} \widetilde{y}\,
\sigma(\widetilde{y}) =
\int_0^{\widetilde{d}/2} {\rm d} \widetilde{x}
\int_{-\widetilde{L}_y/2}^{\widetilde{L}_y/2} {\rm d}
\widetilde{y}\, n(\widetilde{\bf r}) .
\end{equation}

For a boundary with surface charge modulation, an explicit
formula for the local (osmotic) pressure at a point on the boundary was
derived in the PB limit in Ref. \cite{Samaj19}, see equation (2.26) there.
Within the present geometry of symmetrically charged walls with boundaries
at the points $\widetilde{x}=\pm \widetilde{d}/2$, without any dependence
of the relevant quantities on the $z$ axis, this formula simplifies to
\begin{equation} \label{rovnica1}
\beta P(\pm\widetilde{d}/2,\widetilde{y}) =
n(\pm\widetilde{d}/2,\widetilde{y})
- 2\pi\ell_{\rm B} \sigma^2(\widetilde{y}) + \frac{K^2}{8\pi\ell_{\rm B}} 
\left[ \frac{\partial\phi(\pm\widetilde{d}/2,\widetilde{y})}{
\partial\widetilde{y}} \right]^2 .
\end{equation}
The total pressure between the walls is given by averaging the local
pressure over the surface of any of the walls:
\begin{equation} \label{rovnica2}
P(\pm\widetilde{d}/2) = \lim_{\widetilde{L}_y\to\infty} \frac{1}{\widetilde{L}_y}
\int_{-\widetilde{L}_y/2}^{\widetilde{L}_y/2} {\rm d}\widetilde{y}\,
P(\pm\widetilde{d}/2,\widetilde{y}) . 
\end{equation}

A special property of PB theory is that there are infinitely many
possible representations of the local pressure identified by the
coordinate $\widetilde{x}\in [-\widetilde{d}/2,\widetilde{d}/2]$:
\begin{eqnarray}
\beta P(\widetilde{x},\widetilde{y}) & = & n(\widetilde{x},\widetilde{y})
+ \frac{K^2}{8\pi\ell_{\rm B}} \left\{
\left[ \frac{\partial\phi(\widetilde{x},\widetilde{y})}{
\partial\widetilde{y}} \right]^2 -
\left[ \frac{\partial\phi(\widetilde{x},\widetilde{y})}{
\partial\widetilde{x}} \right]^2 \right\} \nonumber \\
& = & \frac{K^2}{2\pi\ell_{\rm B}} \left[ {\rm e}^{\phi}
+ \frac{1}{4} \left( \frac{\partial\phi}{\partial\widetilde{y}} \right)^2 -
\frac{1}{4} \left( \frac{\partial\phi}{\partial\widetilde{x}} \right)^2
\right] .
\end{eqnarray}
The total pressure is given by
\begin{equation}
P(\widetilde{x}) = \lim_{\widetilde{L}_y\to\infty} \frac{1}{\widetilde{L}_y}
\int_{-\widetilde{L}_y/2}^{\widetilde{L}_y/2} {\rm d}\widetilde{y}\,
P(\widetilde{x},\widetilde{y}) . 
\end{equation}
To prove this, we differentiate $\beta P(\widetilde{x},\widetilde{y})$
with respect to $\widetilde{x}$:
\begin{equation}
\frac{\partial \beta P}{\partial \widetilde{x}} = \frac{K^2}{2\pi\ell_{\rm B}}
\left\{ \frac{\partial\phi}{\partial\widetilde{x}} \left[ {\rm e}^{\phi}
- \frac{1}{2} \left( \frac{\partial^2\phi}{\partial\widetilde{x}^2}
+ \frac{\partial^2\phi}{\partial\widetilde{y}^2} \right) \right]
+ \frac{1}{2} \frac{\partial}{\partial\widetilde{y}}
\left( \frac{\partial\phi}{\partial\widetilde{x}}
\frac{\partial\phi}{\partial\widetilde{y}} \right) \right\} .
\end{equation}
Given the 2D Poisson equation for $\phi$ (\ref{Poissonprime}),
the expression in square brackets vanishes and we get
\begin{equation} \label{eqpress}
\frac{\partial \beta P}{\partial\widetilde{x}} = \frac{K^2}{4\pi\ell_{\rm B}}
\frac{\partial}{\partial\widetilde{y}}
\left( \frac{\partial\phi}{\partial\widetilde{x}}
\frac{\partial\phi}{\partial\widetilde{y}} \right)  .
\end{equation}
Next, we integrate both sides of this equation over
$\widetilde{y}\in (-\widetilde{L}_y/2,\widetilde{L}_y/2)$.
Since $P(\widetilde{x},\widetilde{y})$ and its partial derivatives
are continuous, we can change the order of integration and differentiation
with respect $\widetilde{x}$ on the left-hand side of Eq. (\ref{eqpress}) to get
\begin{equation}
\frac{\partial}{\partial\widetilde{x}}\beta P(\widetilde{x})
= \frac{K^2}{4\pi\ell_{\rm B}}
\lim_{\widetilde{L}_y\to\infty} \frac{1}{\widetilde{L}_y}  
\left( \frac{\partial\phi}{\partial\widetilde{x}}
\frac{\partial\phi}{\partial\widetilde{y}}\Big\vert_{\widetilde{y}=\widetilde{L}/2}
- \frac{\partial\phi}{\partial\widetilde{x}}
\frac{\partial\phi}{\partial\widetilde{y}} \Big\vert_{\widetilde{y}=-\widetilde{L}/2}
\right) .
\end{equation}  
The partial derivatives of the electric field are finite, so
\begin{equation}
\frac{\partial}{\partial\widetilde{x}}\beta P(\widetilde{x}) = 0 .
\end{equation}
This means that for every value of the coordinate
$\widetilde{x}\in [-\widetilde{d}/2,\widetilde{d}/2]$, the pressure
$P(\widetilde{x})$ is equal to the original pressure
$P(\pm\widetilde{d}/2)$ defined by Eqs. (\ref{rovnica1}) and (\ref{rovnica2}). 

The midplane between the walls $\widetilde{x}=0$ is particularly interesting,
because the symmetry
$\phi(\widetilde{x},\widetilde{y}) = \phi(-\widetilde{x},\widetilde{y})$
implies that $\partial\phi/\partial\widetilde{x} = 0$ at $\widetilde{x}=0$.
The local pressure is then given by
\begin{equation} \label{P0}
\beta P(0,\widetilde{y}) = n(0,\widetilde{y}) +\frac{K^2}{8\pi\ell_{\rm B}}
\left[ \frac{\partial\phi(0,\widetilde{y})}{\partial\widetilde{y}} \right]^2 .
\end{equation}
We see that the local pressure is determined by the counterion density in
the midplane plus a positive term due to the variation of the electric
potential along the midplane.
This positive contribution can overcome the negative contribution
due to the decrease in particle density in the midplane.
This means that a decrease in the counterion density at $\widetilde{x}=0$
does not necessarily mean a decrease in pressure between the walls.

\renewcommand{\theequation}{3.\arabic{equation}}
\setcounter{equation}{0}

\section{Uniform surface charge density} \label{uniform}
It is useful to briefly summarize the uniform case of surface charge
densities on walls $\sigma(y,z)=\sigma$, see e.g. Ref. \cite{Andelman06}.
The reduced potential $\phi$ depends only on the $x$-coordinate and the 2D
Liouville equation (\ref{Poissonprime}) reduces to the one-dimensional equation
\begin{equation} \label{1DPB}
\frac{d^2 \phi}{d \widetilde{x}^2}  = 2 {\rm e}^{\phi}  
\end{equation}
where $\widetilde{x}\equiv K_0 x$; the subscript ``0'' in $K_0$ means
``related to the uniform case''.
Multiplying this equation by $\phi'\equiv {\rm d}\phi/{\rm d}\widetilde{x}$
and integrating over $\widetilde{x}$ we get the equation
\begin{equation}
\left( \phi' \right)^2 - 4 {\rm e}^{\phi} = {\rm const.}
\end{equation}
Fixing the gauge, say $\phi(0)=0$, the explicit solution to this equation
looks like
\begin{equation} \label{uniformsolution}
\phi(\widetilde{x}) = - 2 \ln \left( \cos\widetilde{x} \right) .
\end{equation}
It automatically satisfies the boundary condition
$\phi'=0$ at $\widetilde{x}=0$. 
The second boundary condition
\begin{equation} \label{surface}
K_0 \frac{\partial\phi}{\partial \widetilde{x}}
\Big\vert_{\widetilde{x}=\widetilde{d}_0/2} = 4\pi\ell_{\rm B}\sigma
\end{equation}
implies the relation for the dimensionless distance
$\widetilde{d}_0 \equiv K_0 d$ 
\begin{equation} \label{Kd}
\widetilde{d}_0 \tan\left( \frac{\widetilde{d}_0}{2} \right) = \bar{d} ,
\end{equation}
where another dimensionless distance $\bar{d}\equiv d/\mu$ is available
in the sense that it does not contain any unknown parameter like $K_0$.
According to (\ref{densityprime}), the particle density is given by 
\begin{equation} \label{PBdensity}
n(\widetilde{x}) = \frac{K_0^2}{2\pi\ell_{\rm B}}
\frac{1}{\cos^2\widetilde{x}} . 
\end{equation}

The effective interaction between two parallel walls at distance $d$
is given by the pressure $P_0(d)$; the case $P_0>0$ corresponds to repulsion
and $P_0<0$ to attraction of the walls.
Within the two-wall formulation (\ref{bcprimeprime}) and
(\ref{neutralityprime}), the contact value theorem
\cite{Henderson78,Henderson79,Blum81,Wennerstrom82} tells us
that for the present system with counterions only the pressure
is determined by the particle density at the wall
contacts $\widetilde{x}=\pm \widetilde{d}_0/2$ as 
\begin{equation} \label{P00}
\beta P_0 = n(\widetilde{d}_0/2) - 2\pi \ell_{\rm B} \sigma^2 .
\end{equation}
This exact relationship is valid not only in the high-temperature limit,
but also for any temperature.
As shown in the previous section, there are infinitely many
possible representations of the PB pressure identified by
the coordinate $\widetilde{x}$:
\begin{equation} \label{P000}
\beta P_0(\widetilde{x}) = n(\widetilde{x}) - \frac{K_0^2}{8\pi\ell_{\rm B}}
\left( \frac{\partial\phi}{\partial\widetilde{x}} \right)^2 .  
\end{equation}
Choosing $\widetilde{x}=0$, we get
\begin{equation}
\beta P_0 = n(0) .
\end{equation}  
Using the formula for the particle density (\ref{PBdensity})
we obtain for the dimensionless pressure
\begin{equation} \label{pressure}
\bar{P}_0 \equiv \frac{\beta P_0}{2\pi\ell_{\rm B}\sigma^2} 
= \left( K_0 \mu \right)^2 = \left( \frac{\widetilde{d}_0}{\bar{d}} \right)^2 .
\end{equation}
Since the pressure is positive, two uniformly charged walls with
the same surface charge density always repel each other in the PB regime.

In the limit $\bar{d}\to 0$, it is possible to expand the left-hand side
of Eq. (\ref{Kd}) in Taylor powers of small $\widetilde{d}_0$ to obtain
a series dependence of $\widetilde{d}_0$ on $\bar{d}$.  
Substituting this dependence into (\ref{pressure}), we obtain
\begin{equation}
\bar{P}_0 = \frac{2}{\bar{d}} - \frac{1}{3} + \frac{2}{45} \bar{d}
+ O\left( \bar{d}^2\right) .
\end{equation}  
In the limit $\bar{d}\to\infty$, Eq. (\ref{Kd}) implies that
$\widetilde{d}_0 \to \pi$ and the large-distance asymptotic of the pressure
reads as
\begin{equation}
\beta P_0 \mathop{\sim}_{\bar{d}\to\infty} \frac{\pi}{2\ell_{\rm B}} \frac{1}{d^2} . 
\end{equation}
Note that this universal form of pressure does not depend on
the amplitude of the surface charge density $\sigma e$.

\renewcommand{\theequation}{4.\arabic{equation}} 
\setcounter{equation}{0}

\section{Infinitesimal perturbation of uniform surface charge density} 
\label{infinite}
Let's look at the solution of the 2D Liouville equation
(\ref{Poissonprime}) by perturbing slightly the uniform solution
(\ref{uniformsolution}):
\begin{equation} \label{infinitesimal}
\phi(\widetilde{x},\widetilde{y}) = - 2 \ln \left( \cos\widetilde{x} \right) 
+ \sum_{j=1}^{\infty} \varepsilon^j
f_j(\widetilde{x},\widetilde{y}) ,
\end{equation}
where $\varepsilon$ is a positive smallness parameter and
$\widetilde{x}, \widetilde{y}$ are dimensionless coordinates
(\ref{coor}) with a normalization parameter $K$ (\ref{K}) depending on
$\varepsilon$, $K\equiv K(\varepsilon)$.
The smallness of $\varepsilon\ll 1$ is related to the assumption that
the variation of the surface charge density around its average value
is weak, $\vert \sigma(y)-\sigma\vert\ll 1$.
The expansion functions exhibit the reflection symmetry of the reduced
potential (\ref{reflection}), i.e.,
\begin{equation}
f_j(\widetilde{x},\widetilde{y})
= f_j(-\widetilde{x},\widetilde{y})
\qquad \mbox{for all $j=1,2,\ldots$.}
\end{equation}
By substituting the ansatz (\ref{infinitesimal}) into the Liouville equation
(\ref{Poissonprime}) and expanding all functions in powers of $\epsilon$,
the linear term in $\epsilon$ vanishes if
\begin{equation} \label{f1}
\frac{\partial^2 f_1}{\partial \widetilde{x}^2}  + 
\frac{\partial^2 f_1}{\partial \widetilde{y}^2} 
- \frac{2}{\cos^2\widetilde{x}} f_1 = 0 ,  
\end{equation}
the quadratic term in $\epsilon$ vanishes if
\begin{equation} \label{f2}
\frac{\partial^2 f_2}{\partial \widetilde{x}^2}  + 
\frac{\partial^2 f_2}{\partial \widetilde{y}^2} -
\frac{2}{\cos^2\widetilde{x}} f_2 = \frac{f_1^2}{\cos^2\widetilde{x}} ,  
\end{equation}
and so on.

We look for the leading order function of Eq. (\ref{f1}) using the
separation of variables:
\begin{equation}
f_1(\widetilde{x},\widetilde{y}) = \varphi(\widetilde{x}) \psi(\widetilde{y}) .
\end{equation}
The functions $\varphi$ and $\psi$ satisfy the ordinary differential equations
\begin{equation}
\frac{1}{\varphi} \frac{d^2\varphi}{d \widetilde{x}^2} = 
b^2 + \frac{2}{\cos^2\widetilde{x}} , \qquad
\frac{1}{\psi} \frac{d^2\psi}{d \widetilde{y}^2} = - b^2 ,
\end{equation}
where $b$ is a real (say positive) parameter.
The solution for $\psi$ has the form
\begin{equation}
\psi = \cos \left( b \widetilde{y} \right) , 
\end{equation}
where the prefactor is set to unity for simplicity.
There are two independent solutions for $\varphi$, 
\begin{equation}
\varphi_{\pm}(\widetilde{x}) = \frac{{\rm e}^{\pm b\widetilde{x}}}{2} 
\left( b \pm \tan\widetilde{x} \right) ,
\end{equation}
and the solution possessing the reflection symmetry
$\widetilde{x}\to -\widetilde{x}$ is given by
$\varphi(\widetilde{x}) = \varphi_+(\widetilde{x}) + \varphi_-(\widetilde{x})$.
The first expansion function is then
\begin{equation} \label{ef1}
f_1(\widetilde{x},\widetilde{y}) = \cos \left( b \widetilde{y} \right) 
\left[ \tan\widetilde{x} \sinh\left( b \widetilde{x} \right)  +
b \cosh\left( b \widetilde{x} \right)  \right] .
\end{equation}

The reflection-symmetric solution of (\ref{f2}) for the lower-order
function $f_2$ is the sum of the homogeneous solution $f_1$
(multiplied by an arbitrary real constant $c$) plus the inhomogeneous part:
\begin{eqnarray}
f_2(\widetilde{x},\widetilde{y};c) & = & c f_1(\widetilde{x},\widetilde{y})
+ \frac{1}{4} \Bigg\{ b^2 \sinh^2(b\widetilde{x}) \nonumber \\
& & + \cos^2(b\widetilde{y}) \left[
\frac{\sinh^2(b\widetilde{x})}{\cos^2\widetilde{x}} 
- b^2 \cosh(2b\widetilde{x}) \right] \Bigg\} .  \label{ef2}
\end{eqnarray}
Note that the inhomogeneous part of the solution cannot be written
in the form of a separation of variables.

It is difficult to continue to higher order functions $f_3$, $f_4$, etc.
because the corresponding differential equations are more difficult
to solve.

\renewcommand{\theequation}{5.\arabic{equation}}
\setcounter{equation}{0}

\section{Surface charge density modulation: Exact solutions}
\label{exact}

\subsection{General solution of the 2D Liouville equation} \label{general}
The 2D Liouville equation (\ref{Poissonprime}) is a partial differential
equation of elliptic type that has many applications in physics.
Several partial exact solutions of this equation have been determined
in the past.
The most general solution \cite{Crowdy97} is expressible in terms
of complex variables
\begin{equation} \label{0}
z = \widetilde{x} + {\rm i}\widetilde{y} , \qquad
\bar{z} = \widetilde{x} - {\rm i}\widetilde{y}
\end{equation}  
as follows
\begin{eqnarray}
\phi(\widetilde{x},\widetilde{y}) & = & - 2 \ln \left[
c_1 Y_1(z) \bar{Y}_1(\bar{z}) + c_2 Y_2(z) \bar{Y}_2(\bar{z})
\right. \nonumber \\ & & \left.
+ c_3 Y_1(z) \bar{Y}_2(\bar{z}) + \bar{c}_3 \bar{Y}_1(\bar{z}) Y_2(z) \right]
+ \ln \left[ W(z) \bar{W}(\bar{z}) \right] , \label{1}
\end{eqnarray}
where the two independent ``trial'' functions $Y_1(z)$ and $Y_2(z)$ are
analytic in $z$ and exhibit nonzero Wronskian
\begin{equation} \label{2}
W(z) \equiv Y_1(z) Y'_2(z) - Y'_1(z) Y_2(z) ,
\end{equation}
$c_1, c_2$ are real constants and $c_3$ a complex constant, constrained by
\begin{equation} \label{3}
\vert c_3\vert^2 - c_1 c_2 = \frac{1}{4} .
\end{equation}  
Here the conjugate function $\bar{f}(\bar{z})$ is defined as
$\bar{f}(\bar{z})\equiv \overline{f(z)}$.

\subsection{Particular solutions for the two-wall geometry} \label{particular}
It is instructive to reproduce the exact result for two uniformly charged
walls at a distance $d$ (\ref{uniformsolution}) with respect to the general
2D solution (\ref{1})-(\ref{3}).
Let us choose the following trial functions
\begin{equation} \label{trial1}
Y_1(z) = {\rm e}^{{\rm i}z/2} , \qquad Y_2(z) = {\rm e}^{-{\rm i}z/2}
\end{equation}
and fix the constants as follows
\begin{equation} \label{constants}
c_1 = c_2 = 0 , \qquad c_3 = \frac{1}{2} .
\end{equation}  
Since
\begin{equation}
\frac{1}{2} \left[ Y_1(z) \bar{Y}_2(\bar{z}) + \bar{Y}_1(\bar{z}) Y_2(z) \right]
= \cos(\widetilde{x})
\end{equation}
and $W(z)=-{\rm i}$, $\bar{W}(\bar{z})={\rm i}$, the solution (\ref{1})
coincides with the uniform one (\ref{uniformsolution}), as intended.

Now let the trial functions be chosen as
\begin{equation} \label{trial3}
Y_1(z) = {\rm e}^{{\rm i}z/2} \left[ 1 + {\rm i} a \sinh( b z) \right] ,
\quad
Y_2(z) = {\rm e}^{-{\rm i}z/2} \left[ 1 - {\rm i} a \sinh( b z) \right] ,
\end{equation}  
where $a, b$ are positive real parameters and the constants $c_1, c_2, c_3$
are chosen as before in Eq. (\ref{constants}).
These trial functions reduce themselves to the previous ones when $a=0$.
After some simple algebra, the explicit solution for $\phi$ is
\begin{equation} \label{fii}
\phi(\widetilde{x},\widetilde{y}) = - 2 \ln \cos\widetilde{x}
- 2 \ln g(\widetilde{x},\widetilde{y}) + \ln h(\widetilde{x},\widetilde{y}) ,
\end{equation}  
where
\begin{equation} \label{defg}  
g(\widetilde{x},\widetilde{y}) = 1 - a^2\cosh^2(b \widetilde{x})
+ a^2 \cos^2(b \widetilde{y}) - 2 a \cos(b \widetilde{y})
\tan \widetilde{x} \sinh(b \widetilde{x})
\end{equation}  
and
\begin{eqnarray}
h(\widetilde{x},\widetilde{y}) & = &
\left[ 1 + 2 a b \cosh(b z) + a^2 \sinh^2(b z) \right] \nonumber \\
& & \qquad \times
\left[ 1 + 2 a b \cosh(b \bar{z}) + a^2 \sinh^2(b \bar{z}) \right]
\nonumber \\ & = &
\Big\{ 1 + 2 a b \cosh(b \widetilde{x}) \cos(b \widetilde{y})
+ 2 a^2 \cosh^2(b \widetilde{x}) \cos^2(b \widetilde{y})
\nonumber \\ &  &
- a^2 \left[ \cosh^2(b \widetilde{x}) + \cos^2(b \widetilde{y}) \right] \Big\}^2
\nonumber \\ &  &
+ 4 a^2 \sinh^2(b \widetilde{x}) \sin^2(b \widetilde{y})
\left[ b + a \cosh(b \widetilde{x}) \cos(b \widetilde{y}) \right]^2 . 
\label{hh}
\end{eqnarray}  
Note that the electrostatic potential $\phi$ has the required
reflection symmetry
$\phi(\widetilde{x},\widetilde{y}) = \phi(-\widetilde{x},\widetilde{y})$.
It is easy to verify, e.g. using the symbolic calculation
program \emph{Mathematica}, that this solution indeed satisfies
the 2D Liouville equation (\ref{Poissonprime}).

The Taylor expansion of the exact $\phi$ (\ref{fii}) in powers of
the parameter $a$ leads to
\begin{equation}
\phi(\widetilde{x},\widetilde{y}) = - 2 \ln \cos\widetilde{x}
+ 4 a f_1(\widetilde{x},\widetilde{y}) + (4 a)^2
f_2(\widetilde{x},\widetilde{y};c=0) + {\cal O}\left( a^3\right) , 
\end{equation}
where $f_1$ and $f_2$ are the expansion functions (\ref{ef1}) and (\ref{ef2})
of the small-$\varepsilon$ expansion of the electrostatic potential
(\ref{infinitesimal}) obtained in the previous section.
We see that $a$ is related to $\varepsilon$ by $a=\varepsilon/4$.
The exact solution (\ref{fii}) -- (\ref{hh}) explicitly includes all
higher-order $f$-functions in the ansatz (\ref{infinitesimal}), which
are difficult to obtain otherwise.

\renewcommand{\theequation}{6.\arabic{equation}}
\setcounter{equation}{0}

\section{Conditions for regular potential solutions} \label{conditions}
The solution for the reduced electrostatic potential (\ref{fii}) must be
well-defined, in particular real and finite (without any singularity),
over the entire interval of possible values of $x$ and over one period along
the $\widetilde{y}$ axis, i.e.
\begin{equation} \label{region}
\widetilde{x}\in [0,\widetilde{d}/2] , \qquad
\widetilde{y}\in [0,2\pi/b) .
\end{equation}
This puts some constraints on the values of the model parameters
$a$, $b$ and $\widetilde{d}$.
Numerical verification of the regularity of the potential
(\ref{fii}) -- (\ref{hh}) is laborious, because it must be performed on
a sufficiently dense network of discrete points covering the entire
region (\ref{region}).
Therefore, we derive all necessary and sufficient conditions for
regularity analytically.

Since $\cos\widetilde{x}$ under the logarithm must be positive for any value
$\widetilde{x}$, the positive value of $\widetilde{d}$ is limited
as follows
\begin{equation} \label{widetilded}
0 \le \widetilde{d} < \pi .
\end{equation}

Further restrictions concern the functions $g$ (\ref{defg}) and $h$ (\ref{hh}),
which are under logarithm and therefore must be positive
in the entire region (\ref{region}).
Next, we present a general analysis of all necessary and sufficient conditions
for the parameters $a$, $b$ and $\widetilde{d}$ that guarantee the positivity
of the functions $g$ and $h$, which is quite complex and is presented in
the following text.

\subsection{$g(\widetilde{x},\widetilde{y})>0$}
Let $(\widetilde{x}_{\rm min},\widetilde{y}_{\rm min})$ denote the minimum
of $g$ given by (\ref{defg}) in the domain (\ref{region}).
The required positivity of $g(\widetilde{x},\widetilde{y})$
is equivalent to the condition
\begin{equation} \label{conditiong}
g(\widetilde{x}_{\rm min},\widetilde{y}_{\rm min}) > 0 .  
\end{equation}

For a fixed value of $\widetilde{x}$, the minimum of the function
$g(\widetilde{x},\widetilde{y})$ on the $\widetilde{y}$ axis is
determined by the conditions
\begin{equation}
\frac{\partial g(\widetilde{x},\widetilde{y})}{\partial\widetilde{y}}
\Big\vert_{\widetilde{y}=\widetilde{y}_{\min}} = 0 , \qquad
\frac{\partial^2 g(\widetilde{x},\widetilde{y})}{\partial\widetilde{y}^2}
\Big\vert_{\widetilde{y}=\widetilde{y}_{\min}} > 0 .
\end{equation}
Since
\begin{equation}
\frac{\partial g(\widetilde{x},\widetilde{y})}{\partial\widetilde{y}}
= 2 a b \sin\left( b\widetilde{y}\right) \left[
\sinh\left( b \widetilde{x} \right) \tan\widetilde{x} -
a \cos\left( b \widetilde{y}\right) \right] , 
\end{equation}
there are three candidates for the minimum point:
\begin{equation} \label{candidates}
\widetilde{y}_1=0 , \qquad \widetilde{y}_2 = \frac{\pi}{b} , \qquad
\cos\left( b\widetilde{y}_3\right) = \frac{\sinh\left( b \widetilde{x} \right)
\tan\widetilde{x}}{a} . 
\end{equation}
The relation
\begin{equation}
\frac{\partial^2 g(\widetilde{x},\widetilde{y})}{\partial\widetilde{y}^2}
\Big\vert_{\widetilde{y}=\widetilde{y}_1} = 2 a b^2 \left[
\sinh\left( b \widetilde{x} \right) \tan\widetilde{x} - a \right]
\end{equation}  
tells us that $\widetilde{y}_1$ is a minimum point if
\begin{equation} \label{cond1}
a<\sinh\left( b \widetilde{x} \right) \tan\widetilde{x} .
\end{equation}
The relation
\begin{equation}
\frac{\partial^2 g(\widetilde{x},\widetilde{y})}{\partial\widetilde{y}^2}
\Big\vert_{\widetilde{y}=\widetilde{y}_2} = - 2 a b^2 \left[
\sinh\left( b \widetilde{x} \right) \tan\widetilde{x} + a \right]
\end{equation}
tells us that $\widetilde{y}_2$ is always a maximum point.
Finally, the relation
\begin{equation}
\frac{\partial^2 g(\widetilde{x},\widetilde{y})}{\partial\widetilde{y}^2}
\Big\vert_{\widetilde{y}=\widetilde{y}_3} = 2 b^2 \left[
a - \sinh\left( b \widetilde{x} \right) \tan\widetilde{x} \right]
\left[ \sinh\left( b \widetilde{x} \right) \tan\widetilde{x} + a \right]
\end{equation}
tells us that $\widetilde{y}_3$ is a minimum point if
\begin{equation} \label{cond2}
a > \sinh\left( b \widetilde{x} \right) \tan\widetilde{x} .
\end{equation}

Let's start with $\widetilde{y}_1=0$ which is the minimum point if
$a<\sinh\left( b \widetilde{x} \right) \tan\widetilde{x}$.
The corresponding function
\begin{equation}
g(\widetilde{x},0) = 1 - a^2 \sinh^2\left( b \widetilde{x} \right)
- 2 a \sinh\left( b \widetilde{x} \right) \tan\widetilde{x}
\end{equation}  
has a negative derivative with respect to $\widetilde{x}$:
\begin{eqnarray}
\frac{{\rm d} g(\widetilde{x},0)}{{\rm d} \widetilde{x}}
& = & - 2 a^2 b \sinh(b\widetilde{x}) \cosh(b\widetilde{x}) \nonumber \\
& & -2 a \left[ \frac{\sinh(b\widetilde{x})}{\cos^2\widetilde{x}}
+ b \cosh(b\widetilde{x}) \tan\widetilde{x} \right] < 0 .  
\end{eqnarray}  
This means that the minimum of $g(\widetilde{x},0)$ is reached at the endpoint
$\widetilde{x}=\widetilde{d}/2$, where the condition (\ref{cond1})
takes the form
\begin{equation} \label{cond3}
a<\sinh(b \widetilde{d}/2) \tan(\widetilde{d}/2) .
\end{equation}  
The positivity requirement $g(\widetilde{d}/2,0)>0$ is equivalent to
the inequality
\begin{equation}
1 - a^2 \sinh^2(b\widetilde{d}/2) - 2 a \sinh(b\widetilde{d}/2)
\tan(\widetilde{d}/2) > 0 . 
\end{equation}  
From this inequality, it follows that
\begin{equation} \label{cond4}
0\le a < \frac{1}{\sinh(b \widetilde{d}/2)}
\frac{1-\sin(\widetilde{d}/2)}{\cos(\widetilde{d}/2)} .
\end{equation}  
Given conditions (\ref{cond3}) and (\ref{cond4}),
the parameter $a$ is limited by the relation
\begin{equation} \label{equation1}
0 \le a < \min \left\{
\sinh(b \widetilde{d}/2) \tan(\widetilde{d}/2),
\frac{1}{\sinh(b \widetilde{d}/2)}
\frac{1-\sin(\widetilde{d}/2)}{\cos(\widetilde{d}/2)} \right\} .
\end{equation}  

Let's continue with $\widetilde{y}_3$ given by (\ref{candidates})
which is the minimum point if
$a>\sinh\left( b \widetilde{x} \right) \tan\widetilde{x}$.
We need to find the minimum of the function  
\begin{equation} \label{G1}
g(\widetilde{x},y_3)
= 1 - a^2 \cosh^2(b\widetilde{x}) - \sinh^2(b\widetilde{x}) 
\tan^2\widetilde{x} 
\end{equation}
on the interval $\widetilde{x}\in [0,\widetilde{d}/2]$.
Since
\begin{eqnarray}
\frac{{\rm d}g(\widetilde{x},y_3)}{{\rm d}\widetilde{x}}
& = & - 2 a^2 b \cosh(b\widetilde{x}) \sinh(b\widetilde{x})
\nonumber \\ & &
- 2 \tan\widetilde{x} \sinh(b\widetilde{x})
\left[ \frac{\sinh(b\widetilde{x})}{\cos^2\widetilde{x}}
+ b \tan\widetilde{x} \cosh(b\widetilde{x}) \right] < 0 , 
\end{eqnarray}  
the minimum of $g(\widetilde{x},y_3)$ is reached at the endpoint
$\widetilde{x} = \widetilde{d}/2$
where the condition (\ref{cond2}) reads as
\begin{equation} \label{cond5}
a>\sinh(b \widetilde{d}/2) \tan(\widetilde{d}/2) .
\end{equation}
and the condition (\ref{conditiong}) gives
\begin{equation} \label{firstcond}
1 - a^2 \cosh^2(b\widetilde{d}/2) - \tan^2(\widetilde{d}/2) 
\sinh^2(b\widetilde{d}/2) > 0 .
\end{equation}  
This inequality means that
\begin{equation}
a < \frac{\sqrt{1-\sinh^2(b \widetilde{d}/2)\tan^2(\widetilde{d}/2)}}{
\cosh(b\widetilde{d}/2)} .  
\end{equation}
Finally, the parameter $a$ is now constrained by the relation
\begin{equation} \label{restriction}
\sinh(b \widetilde{d}/2) \tan(\widetilde{d}/2) \le a <
\frac{\sqrt{1-\sinh^2(b \widetilde{d}/2)\tan^2(\widetilde{d}/2)}}{
\cosh(b\widetilde{d}/2)} .  
\end{equation}  

In order to have a non-empty set of $a$-values determined by
(\ref{restriction}), it is necessary that the lower bound is smaller
than the upper bound, i.e.
\begin{equation} \label{ineq1}
\sinh(b \widetilde{d}/2) \tan(\widetilde{d}/2) <
\frac{\sqrt{1-\sinh^2(b \widetilde{d}/2)\tan^2(\widetilde{d}/2)}}{
\cosh(b\widetilde{d}/2)} .  
\end{equation}
This inequality is equivalent to the equation
\begin{equation}
\tan^2(\widetilde{d}/2) \sinh^4(b \widetilde{d}/2) +
2 \tan^2(\widetilde{d}/2) \sinh^2(b \widetilde{d}/2) - 1 < 0 ,   
\end{equation}  
which leads to
\begin{equation} \label{ineq2}
\sinh(b \widetilde{d}/2) \tan(\widetilde{d}/2) <
\frac{1}{\sinh(b \widetilde{d}/2)}
\frac{1-\sin(\widetilde{d}/2)}{\cos(\widetilde{d}/2)} .
\end{equation}
Combining the relations (\ref{equation1}) and (\ref{restriction}),
we conclude that if the inequality (\ref{ineq2}) holds, the values of $a$
are restricted to
\begin{equation}
0\le a < \frac{\sqrt{1-\sinh^2(b \widetilde{d}/2)\tan^2(\widetilde{d}/2)}}{
\cosh(b\widetilde{d}/2)} .  
\end{equation}  

Now consider the inequality opposite to that in (\ref{ineq1}), i.e.
\begin{equation} \label{ineq3}
\sinh(b \widetilde{d}/2) \tan(\widetilde{d}/2) >
\frac{\sqrt{1-\sinh^2(b \widetilde{d}/2)\tan^2(\widetilde{d}/2)}}{
\cosh(b\widetilde{d}/2)} .
\end{equation}
This inequality can be shown to be equivalent to the one 
\begin{equation} \label{ineq4}
\sinh(b \widetilde{d}/2) \tan(\widetilde{d}/2) >
\frac{1}{\sinh(b \widetilde{d}/2)}
\frac{1-\sin(\widetilde{d}/2)}{\cos(\widetilde{d}/2)} .
\end{equation}
Then the restriction (\ref{restriction}) does not provide any solution
for $a$ and by the restriction (\ref{equation1}) the values of
$a$ are limited to the interval
\begin{equation} \label{equation2}
0 \le a < \frac{1}{\sinh(b \widetilde{d}/2)}
\frac{1-\sin(\widetilde{d}/2)}{\cos(\widetilde{d}/2)} .
\end{equation}  

\subsection{$h(\widetilde{x},\widetilde{y})>0$}
The function $h(\widetilde{x},\widetilde{y})$ (\ref{hh}) is the sum of
two squares and as such can be either positive or equal to 0.
The ``dangerous'' equality $h(\widetilde{x},\widetilde{y})=0$ is
satisfied if the two equations
\begin{eqnarray}
\sinh(b\widetilde{x}) \sin(b\widetilde{y}) \left[ b + a \cosh(b\widetilde{x})
\cos(b\widetilde{y})  \right] & = & 0 , \label{eq1} \\
1 + 2 a b \cosh(b\widetilde{x}) \cos(b\widetilde{y})
+ 2 a^2 \cosh^2(b\widetilde{x}) \cos^2(b\widetilde{y}) & & \nonumber \\
- a^2 \left[ \cosh^2(b\widetilde{x}) + \cos^2(b\widetilde{y}) \right] & = & 0,
\label{eq2}
\end{eqnarray}
are valid simultaneously.
Eq. (\ref{eq1}) is satisfied for $\widetilde{x}=0$, or $\widetilde{y}=0$,
or $\widetilde{y}=\pi/b$, or $b+a\cosh(b\widetilde{x})\cos(b\widetilde{y})=0$.

\noindent $\bullet$ $\widetilde{x}=0$:
In this case the second equation (\ref{eq2}) reads
\begin{equation}
a^2 \cos^2(b\widetilde{y}) + 2 a b \cos(b\widetilde{y}) + (1-a^2) = 0 .
\end{equation}  
Its solution is
\begin{equation} \label{solution}
\cos(b\widetilde{y}) = - \frac{b}{a} \pm \frac{1}{a} \sqrt{a^2+b^2-1} .
\end{equation}  
There is no real solution for $\cos(b\widetilde{y})$ provided that
\begin{equation} \label{secondrestriction}
a^2 + b^2 < 1 .
\end{equation}
Since $-1\le \cos(b\widetilde{y}) \le 1$, a real solution for
$\widetilde{y}$ is missing also if
\begin{equation} \label{restr}
a^2 + b^2 = 1 , \qquad b>a .
\end{equation}  
When $a^2+b^2>1$, a real solution for $\cos(b\widetilde{y})$
(\ref{solution}) must lie outside the interval $[-1,1]$ to
ensure the positivity of $h$.
In particular, if we put a $(-)$ sign in front of the square root on
the right-hand side of (\ref{solution}), we have
\begin{equation}
- \frac{b}{a} - \frac{1}{a} \sqrt{a^2+b^2-1} < -1 ,
\end{equation}
or equivalently,
\begin{equation}
a-b < \sqrt{a^2+b^2-1} . 
\end{equation}  
This inequality is definitely true if $a<b$.
When $a>b$, bringing each side of the inequality to the power of 2
gives us $2 a b >1$.
If we take the sign $(+)$ before the square root on the right-hand side
of (\ref{solution}), we must distinguish between two cases:
$a^2>1$. when the inequality
\begin{equation}
- \frac{b}{a} + \frac{1}{a} \sqrt{a^2+b^2-1} > 1 
\end{equation}
is equivalent to the requirement $2 a b < -1$, which has no solution
for positive parameters $a$ and $b$, and $a^2<1$, when the inequality
\begin{equation}
- \frac{b}{a} + \frac{1}{a} \sqrt{a^2+b^2-1} < -1 
\end{equation}
has a solution if $a<b$ and $2 a b <1$. 
To conclude this part, $h>0$ for $\widetilde{x}=0$ if one of the conditions
(\ref{secondrestriction}) or (\ref{restr}) holds, or if the conditions
\begin{equation}
a^2+b^2>1 , \qquad a^2<1, \qquad a<b , \qquad 2 a b<1
\end{equation}  
hold simultaneously. 

\noindent $\bullet$ $\widetilde{y}=0$:
In this case the second equation (\ref{eq2}) has the form
\begin{equation}
a^2 \sinh^2(b\widetilde{x})  + 2 a b \cosh(b\widetilde{x}) + 1 = 0 .
\end{equation}
The function on the left-hand side is always positive for
$\widetilde{x}\in [0,\widetilde{d}/2]$.
Consequently, $h$ is always positive for $\widetilde{y}=0$, which
does not impose any restrictions on the possible values of
the parameters $a$, $b$ and $\widetilde{d}$.

\noindent $\bullet$ $\widetilde{y}=\pi/b$:
In this case, the second equation (\ref{eq2}) reads
\begin{equation}
a^2 \cosh^2(b\widetilde{x}) - 2 a b \cosh(b\widetilde{x}) + (1-a^2) = 0 .
\end{equation}  
Its solution is
\begin{equation} \label{solution2}
\cosh(b\widetilde{x}) = \frac{b}{a} \pm \frac{1}{a} \sqrt{a^2+b^2-1} .
\end{equation}  
There is no real solution for $\cosh(b\widetilde{x})$ assuming that
\begin{equation} \label{secondrestriction2}
a^2 + b^2 < 1 .
\end{equation}
Eq. (\ref{solution2}) has no solution for $\widetilde{x}$ if $a^2+b^2=1$
and at the same time $b<a$.
If $a^2+b^2>1$, Eq. (\ref{solution2}) has no solution for $\widetilde{x}$
in two cases:
\begin{equation}
\frac{b}{a} + \frac{1}{a} \sqrt{a^2+b^2-1} <1 , \qquad
\frac{b}{a} - \frac{1}{a} \sqrt{a^2+b^2-1} > \cosh(b\widetilde{d}/2) .  
\end{equation}  

\noindent $\bullet$ $b+a\cosh(b\widetilde{x})\cos(b\widetilde{y})=0$:
This condition is equivalent to the condition
\begin{equation} \label{eqforcos}
-\cos(b\widetilde{y}) = \frac{b}{a\cosh(b\widetilde{x})} .
\end{equation}  
The function $1/\cosh(b\widetilde{x})$ decreases with increasing
$\widetilde{x}$, so if
\begin{equation}
a < \frac{b}{\cosh(b\widetilde{d}/2)}
\end{equation}  
there is no solution for $-\cos(b\widetilde{y})$ in the interval $[0,1]$.
On the other hand, if
\begin{equation}
a > \frac{b}{\cosh(b\widetilde{d}/2)} ,
\end{equation}
substituting $\cos(b\widetilde{y})$ from (\ref{eqforcos}) into
the second equation (\ref{eq2}) yields
\begin{equation} \label{coshcos}
\cosh(b\widetilde{x}) = \frac{\sqrt{1+2ab}+\sqrt{1-2ab}}{2 a} , \qquad  
- \cos(b\widetilde{y}) = \frac{\sqrt{1+2ab}-\sqrt{1-2ab}}{2 a} .   
\end{equation}  
Real solutions for $\widetilde{x}$ and $\widetilde{y}$ do not exist if
\begin{equation}
2 a b > 1 .
\end{equation}  
On the other hand, if
\begin{equation}
2 a b < 1,
\end{equation}
we look for values of parameters $a$ and $b$ such that either
$\cosh(b\widetilde{x})$ given by (\ref{coshcos}) lies outside the interval
\begin{equation}
1 \le \cosh(b\widetilde{x}) \le \cosh(b\widetilde{d}/2) ,
\end{equation}
what happens if
\begin{equation}
0\le a < \frac{1}{\cosh(b\widetilde{d}/2)}
\sqrt{1 - \frac{b^2}{\cosh^2(b\widetilde{d}/2)}} ,  
\end{equation}
or $-\cos(b\widetilde{y})$ given by (\ref{coshcos}) lies outside the interval
\begin{equation}
0 \le -\cos(b\widetilde{y}) \le 1 ,
\end{equation}  
what happens if $a^2+b^2>1$.

\renewcommand{\theequation}{7.\arabic{equation}}
\setcounter{equation}{0}

\section{Physical model generated from potential} \label{analysis}

\subsection{Derivation of physical quantities}
The electric potential defined by equations (\ref{fii}), (\ref{defg})
and (\ref{hh}) is periodic along the $y$ axis with a period equal
to $2\pi/b$.
According to the relation (\ref{bcprimeprime}), the surface charge density
on the wall surface is given by
\begin{equation}
\sigma(\widetilde{y}) = \frac{K}{4\pi\ell_{\rm B}}  
\frac{\partial\phi(\widetilde{x},\widetilde{y})}{
\partial \widetilde{x}}\Big\vert_{\widetilde{x}=\widetilde{d}/2} .
\end{equation}
It is also periodic along the $y$ axis with period $2\pi/b$.
The mean value of the surface charge density is then given by  
\begin{equation} \label{surfcharge}
\langle \sigma \rangle = \frac{K}{4\pi\ell_{\rm B}} \frac{b}{2\pi}  
\int_0^{2\pi/b} {\rm d}\widetilde{y}\,
\frac{\partial\phi(\widetilde{x},\widetilde{y})}{
\partial \widetilde{x}}\Big\vert_{\widetilde{x}=\widetilde{d}/2} .
\end{equation}

The particle density (\ref{densityprime}) and local pressure
(\ref{P0}) are also periodic along the $y$ axis with period $2\pi/b$,
so the mean pressure acting on either of the walls is given by
\begin{equation} \label{P0000}
\beta \langle P \rangle = \frac{K^2}{2\pi\ell_{\rm B}} \frac{b}{2\pi}
\int_0^{2\pi/b} {\rm d}\widetilde{y}\, \left\{
{\rm e}^{\phi(0,\widetilde{y})} + \frac{1}{4}
\left[\frac{\partial\phi(0,\widetilde{y})}{\partial\widetilde{y}} \right]^2
\right\} .
\end{equation}
The dimensionless pressure has the form
\begin{equation} \label{newpressure}
\bar{P} \equiv \frac{\beta \langle P\rangle}{2\pi
\ell_{\rm B}\langle\sigma\rangle^2} = \frac{8\pi}{b}
\frac{\int_0^{2\pi/b} {\rm d}\widetilde{y}\, \left\{
{\rm e}^{\phi(0,\widetilde{y})} + \frac{1}{4}
\left[\frac{\partial\phi(0,\widetilde{y})}{\partial\widetilde{y}} \right]^2
\right\}}{\left[ \int_0^{2\pi/b} {\rm d}\widetilde{y}\,
\frac{\partial\phi(\widetilde{x},\widetilde{y})}{
\partial \widetilde{x}}\Big\vert_{\widetilde{x}=\widetilde{d}/2}\right]^2} .
\end{equation}

The relationship between the dimensionless distances $\widetilde{d}$ and
$\bar{d}$ follows from the definition of the latter:
\begin{equation}
\bar{d} \equiv 2\pi \ell_{\rm B} \langle\sigma\rangle d
= \widetilde{d} \frac{b}{4\pi} \int_0^{2\pi/b} {\rm d}\widetilde{y}\,
\frac{\partial\phi(\widetilde{x},\widetilde{y})}{
\partial \widetilde{x}}\Big\vert_{\widetilde{x}=\widetilde{d}/2} . 
\end{equation}  
For given values of the parameters $a$, $b$ and $\widetilde{d}$,
this $\bar{d}$ is used to calculate the relevant quantities for the case of
uniform surface charge densities, namely $\widetilde{d}_0$ given by
the relation (\ref{Kd}) and subsequently $\bar{P}_0$ given by
the relation (\ref{pressure}).

\subsection{Results}
The dimensionless pressure and modulated surface charge density
of our exactly solvable model depend on the dimensionless
distance between the walls $\widetilde{d}$ constrained by (\ref{widetilded}),
the amplitude parameter $a$, and the period parameter $b$.

\begin{figure}[]
\begin{center}
\includegraphics[clip,width=0.9\textwidth]{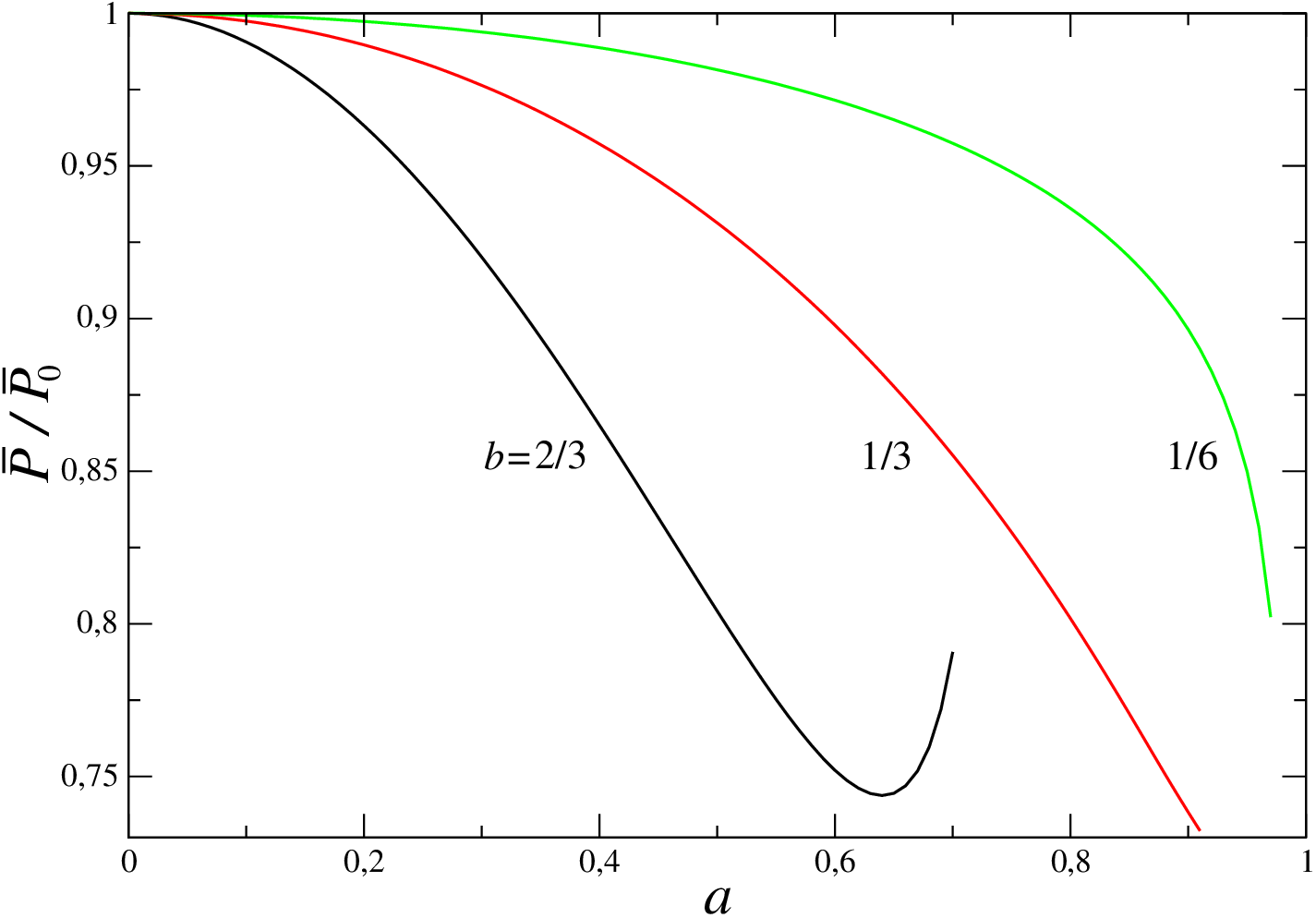}
\caption{Dependence of the ratio $\bar{P}/\bar{P}_0$ on the amplitude
parameter $a$ for the dimensionless distance between the walls
$\widetilde{d}=\pi/2$ and the values of the period parameter
$b=2/3$, $1/3$, $1/6$.}
\label{fig2}
\end{center}
\end{figure}

First, let us consider a relatively large distance $\widetilde{d}=\pi/2$.
The results for the ratio of the dimensionless pressure $\bar{P}$ for
the model with surface charge modulation and the pressure $\bar{P}_0$
for the model with a uniform surface charge density equal to the mean value
of the modulated surface charge density $\langle\sigma\rangle$ given by
the relation (\ref{surfcharge}) (see section \ref{uniform}) are shown in
Fig. \ref{fig2}.
Specifically, for three values of the period parameter $b=2/3$, $1/3$ and
$1/6$, the dependence of the ratio $\bar{P}/\bar{P}_0$ on the amplitude
parameter $a$ is shown by solid lines.
The intervals of allowed values of $a$ are determined by the conditions for
a regular electric potential derived in section \ref{conditions}.
It can be seen that for each value of $b$ the ratio $\bar{P}/\bar{P}_0$ is
less than 1, which means that the modulation of the surface charge density
weakens the pressure between the walls.
Interestingly, for $b=2/3$ the plot of $\bar{P}/\bar{P}_0$
as a function of $a$ is not monotonic.

\begin{figure}[]
\begin{center}
\includegraphics[clip,width=0.9\textwidth]{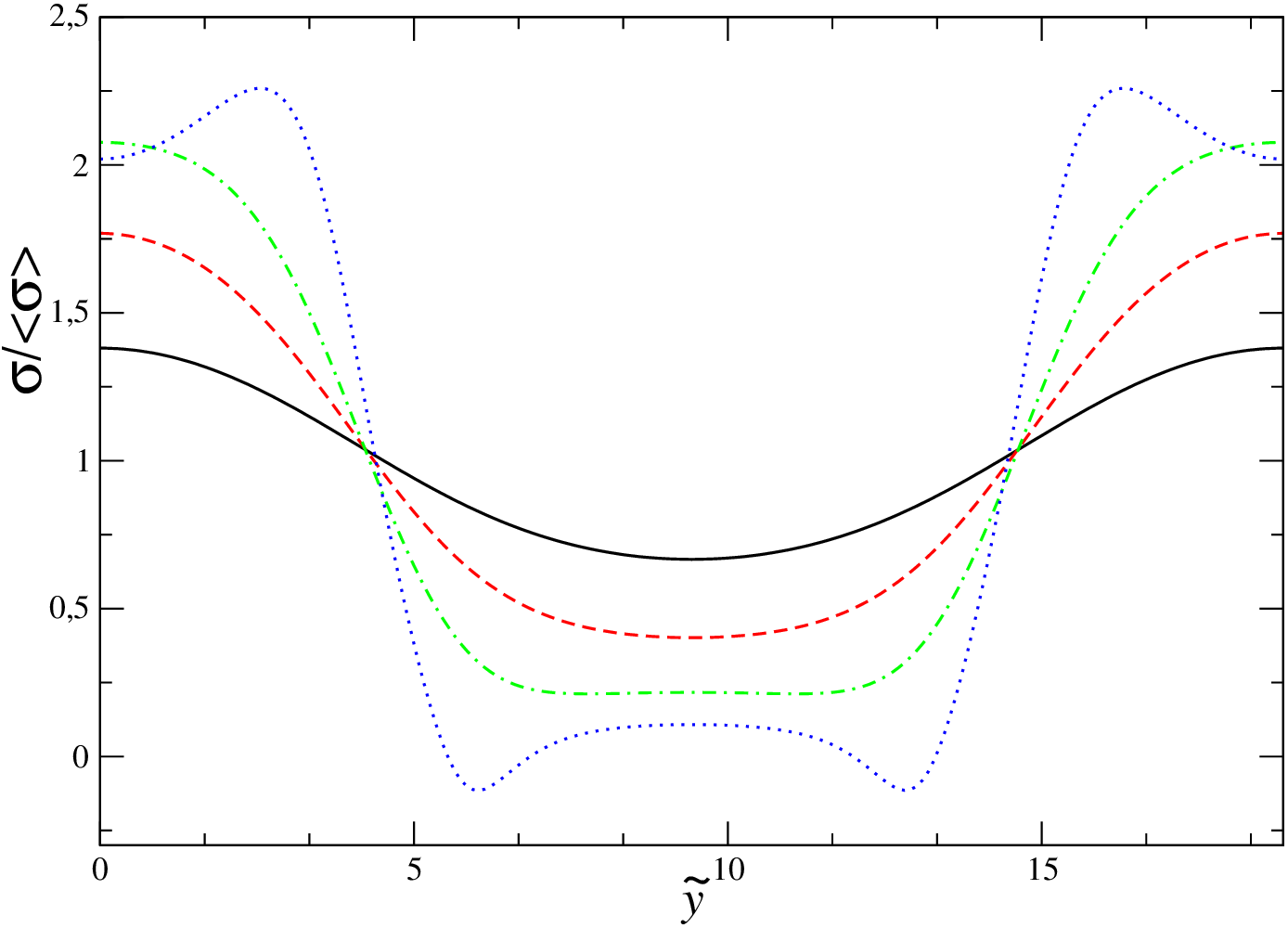}
\caption{Plots of the normalized modulated surface charge density
$\sigma/\langle\sigma\rangle$ on the dimensionless coordinate $\widetilde{y}$
for the dimensionless distance between the walls $\widetilde{d}=\pi/2$
and the period parameter $b=1/3$, plotted within one period $2\pi/b=6\pi$.  
The solid line corresponds to the amplitude parameter $a=1/5$,
the dashed line to $a=2/5$, the dash-dotted line to $a=3/5$, and
the dotted line to $a=4/5$.}
\label{fig3}
\end{center}
\end{figure}

The plots of the normalized modulated surface charge density
$\sigma/\langle\sigma\rangle$ on the dimensionless coordinate $\widetilde{y}$
for the dimensionless wall distance $\widetilde{d}=\pi/2$ and
the period parameter $b=1/3$ are shown for one period $2\pi/b = 6\pi$
in Fig. \ref{fig3}.
The plots become more pronounced with increasing value of $a$, as expected.
The solid line $(a=1/5)$ corresponds to a slowly varying
surface charge density, the change is more pronounced for $a=2/5$
(dashed line) and $a=3/5$ (dash-dotted line), but the surface charge
is always positive with a minimum at $\widetilde{y}=\pi/b=3\pi$.
For a large value of the amplitude parameter $a=4/5$ (dotted line),
the surface charge density exhibits a local maximum at $\widetilde{y}=3\pi$
and is negative at certain intervals of $\widetilde{y}$; in such a case
the term ``counterion'' does not make sense.

\begin{figure}[]
\begin{center}
\includegraphics[clip,width=0.9\textwidth]{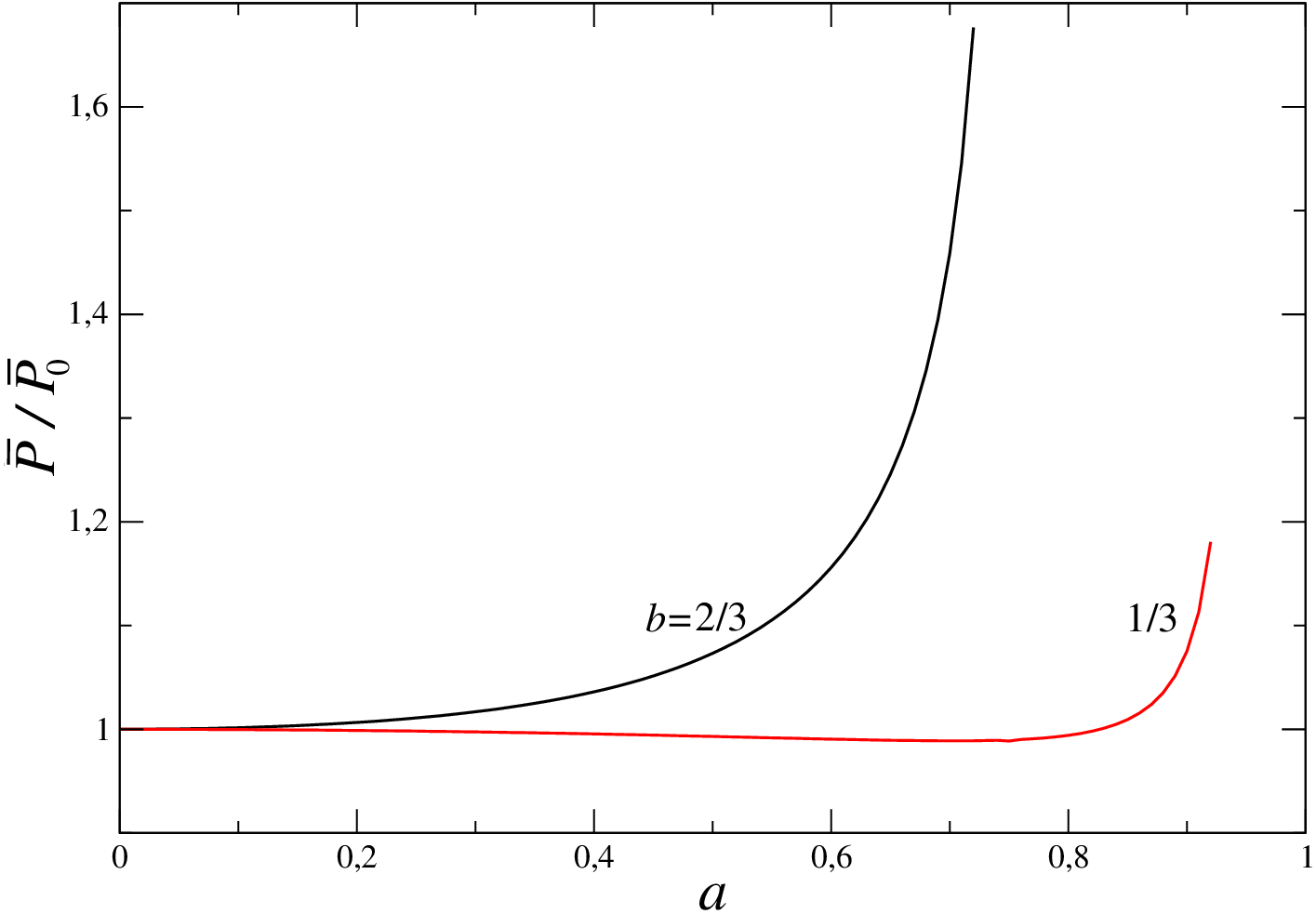}
\caption{Dependence of the ratio $\bar{P}/\bar{P}_0$ on the amplitude
parameter $a$ for the dimensionless distance between the walls
$\widetilde{d}=\pi/4$ and the values of the period parameter
$b=2/3$, $1/3$.}
\label{fig4}
\end{center}
\end{figure}

Let us continue with the results obtained for the smaller distance
$\widetilde{d}=\pi/4$.
The ratio of dimensionless pressures with the same mean surface
charge densities $\bar{P}/\bar{P}_0$ is shown as a function of
the amplitude parameter $a$ for two values of the period parameter
$b=2/3$ and $1/3$ in Fig. \ref{fig4}.
For $b=2/3$ the ratio $\bar{P}/\bar{P}_0$ increases monotonically from 1 at
$a=0$ to 1.6766 at the endpoint $a=0.72$, i.e. modulating surface
charge density surprisingly enhances the force between walls.
For $b=1/3$, the ratio $\bar{P}/\bar{P}_0$ decreases very slowly from
1 at $a=0$ to 0.988 at $a=0.71$ and then increases to 1.181 at the
endpoint $a=0.92$.

\begin{figure}[]
\begin{center}
\includegraphics[clip,width=0.9\textwidth]{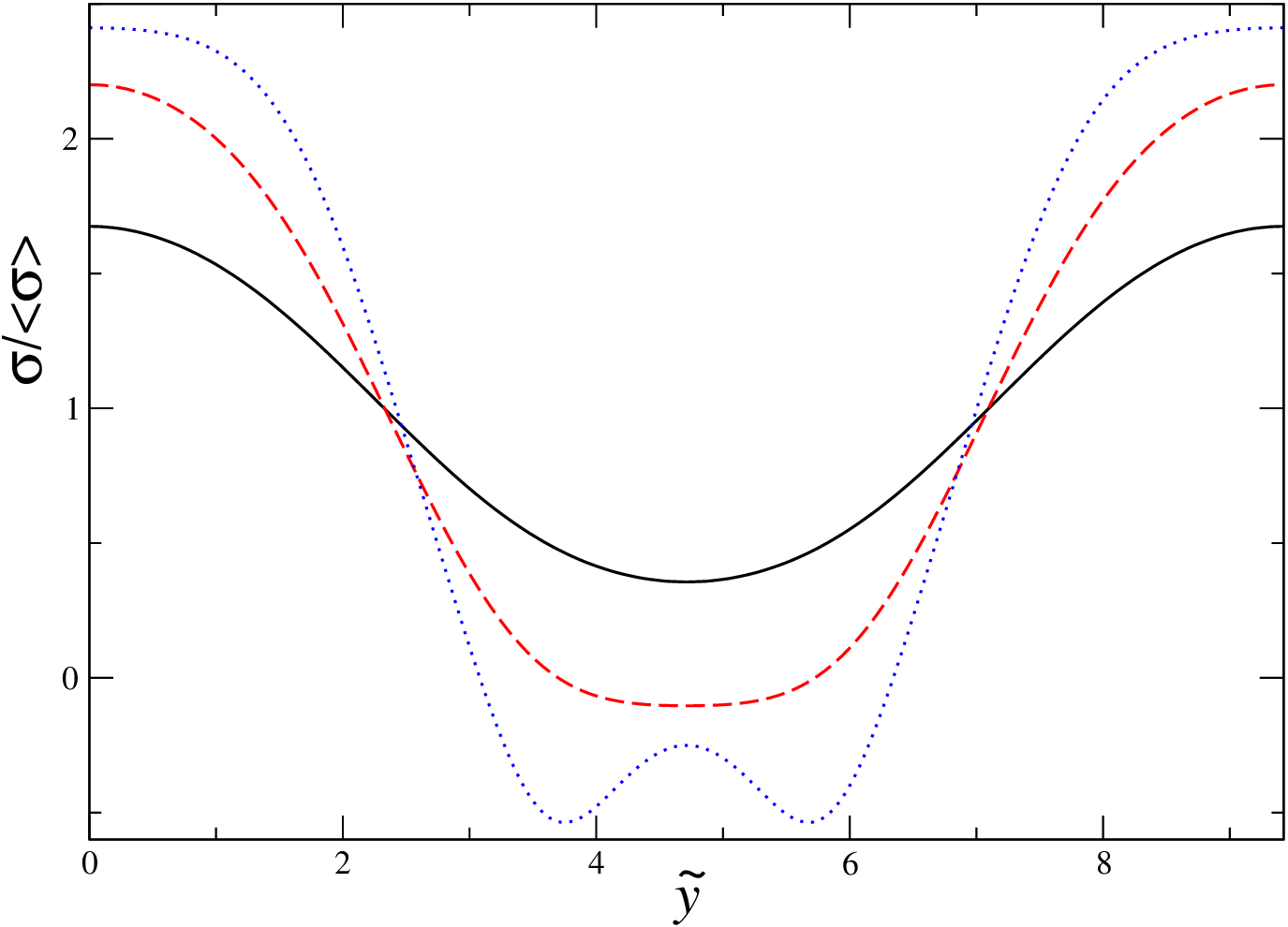}
\caption{Plots of the normalized modulated surface charge density
$\sigma/\langle\sigma\rangle$ on the dimensionless coordinate $\widetilde{y}$
for the dimensionless wall distance $\widetilde{d}=\pi/4$
and the period parameter $b=2/3$, plotted within one period $2\pi/b=3\pi$.  
The solid line corresponds to the amplitude parameter $a=1/5$,
the dashed line to $a=2/5$ and the dotted line to $a=3/5$.}
\label{fig5}
\end{center}
\end{figure}

For $\widetilde{d}=\pi/4$ and $b=2/3$, the profiles of the normalized modulated
surface charge density $\sigma/\langle\sigma\rangle$ as functions
of the coordinate $\widetilde{y}$ for one period $2\pi/b = 3\pi$ are shown 
in Fig. \ref{fig5}.
For the amplitude parameter $a=1/5$ (solid line), the surface charge density
is always positive and exhibits a minimum at $\widetilde{y}=\pi/b=3\pi/2$.
We see that there is an increase in pressure due to
modulation of the surface charge density even for the ``physical''
counterion model with charge distribution $\sigma(\tilde{y})\ge 0$.  
For $a=2/5$ (dashed line), the surface charge density is negative in
a short interval around the minimum point $\widetilde{y}=3\pi/2$.
For $a=3/5$ (dotted line), the surface charge density is negative in
an interval around the local maximum point $\widetilde{y}=3\pi/2$
and a pair of conjugate minimum points.

\begin{figure}[]
\begin{center}
\includegraphics[clip,width=0.9\textwidth]{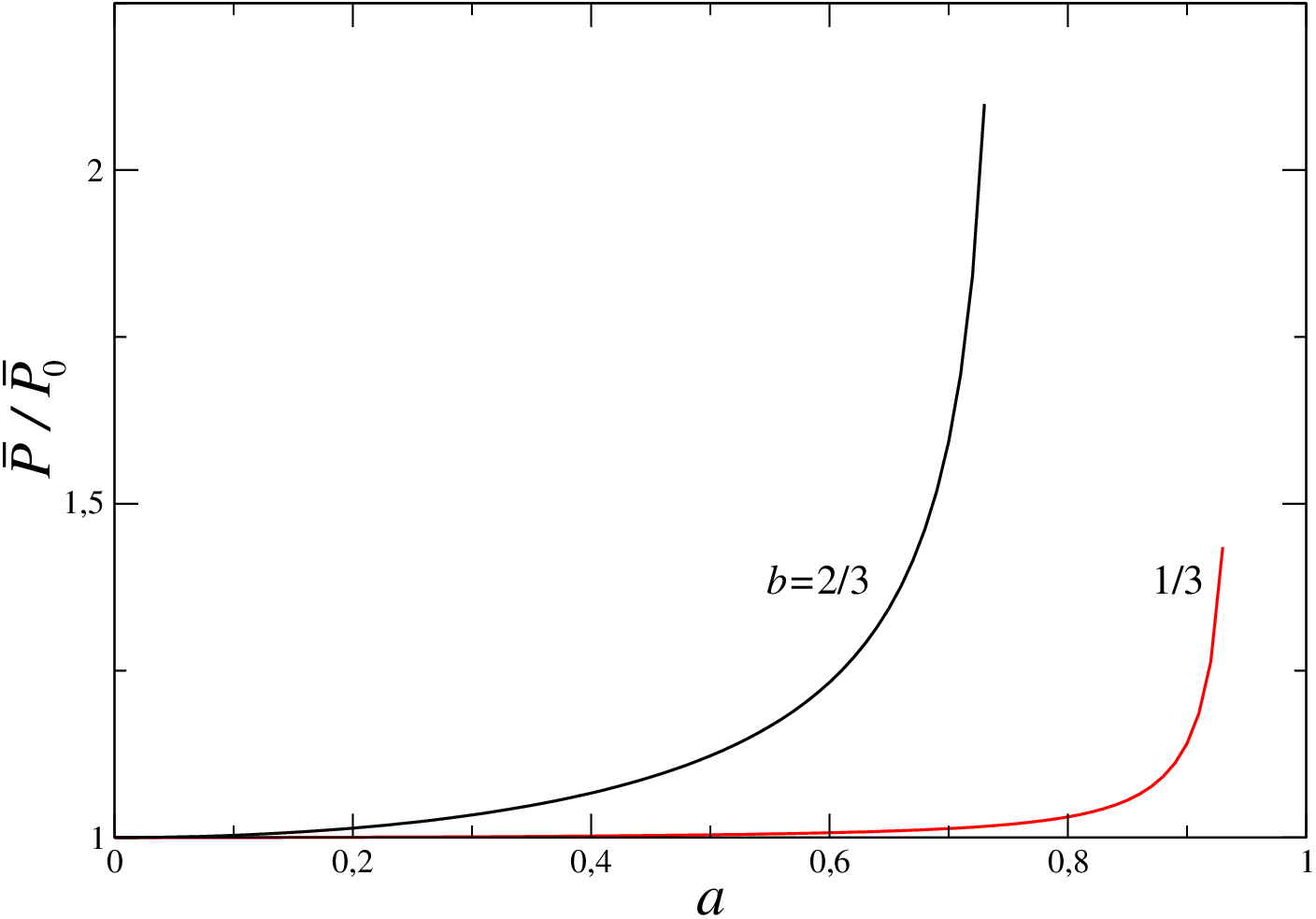}
\caption{
Dependence of the ratio $\bar{P}/\bar{P}_0$ on the amplitude
parameter $a$ for the dimensionless wall distance
$\widetilde{d}=\pi/8$ and values of the period parameter
$b=2/3$, $1/3$.}
\label{fig6}
\end{center}
\end{figure}

The results for a small distance between the walls $\widetilde{d}=\pi/8$
are quite similar to the results for $\widetilde{d}=\pi/4$.
The ratio of dimensionless pressures with the same mean surface
charge densities $\bar{P}/\bar{P}_0$ is shown as a function of
the amplitude parameter $a$ for two values of the period parameter
$b=2/3$ and $1/3$ in Fig. \ref{fig6}.
The ratio is greater than 1 for every allowed value of $a$.

\begin{figure}[]
\begin{center}
\includegraphics[clip,width=0.9\textwidth]{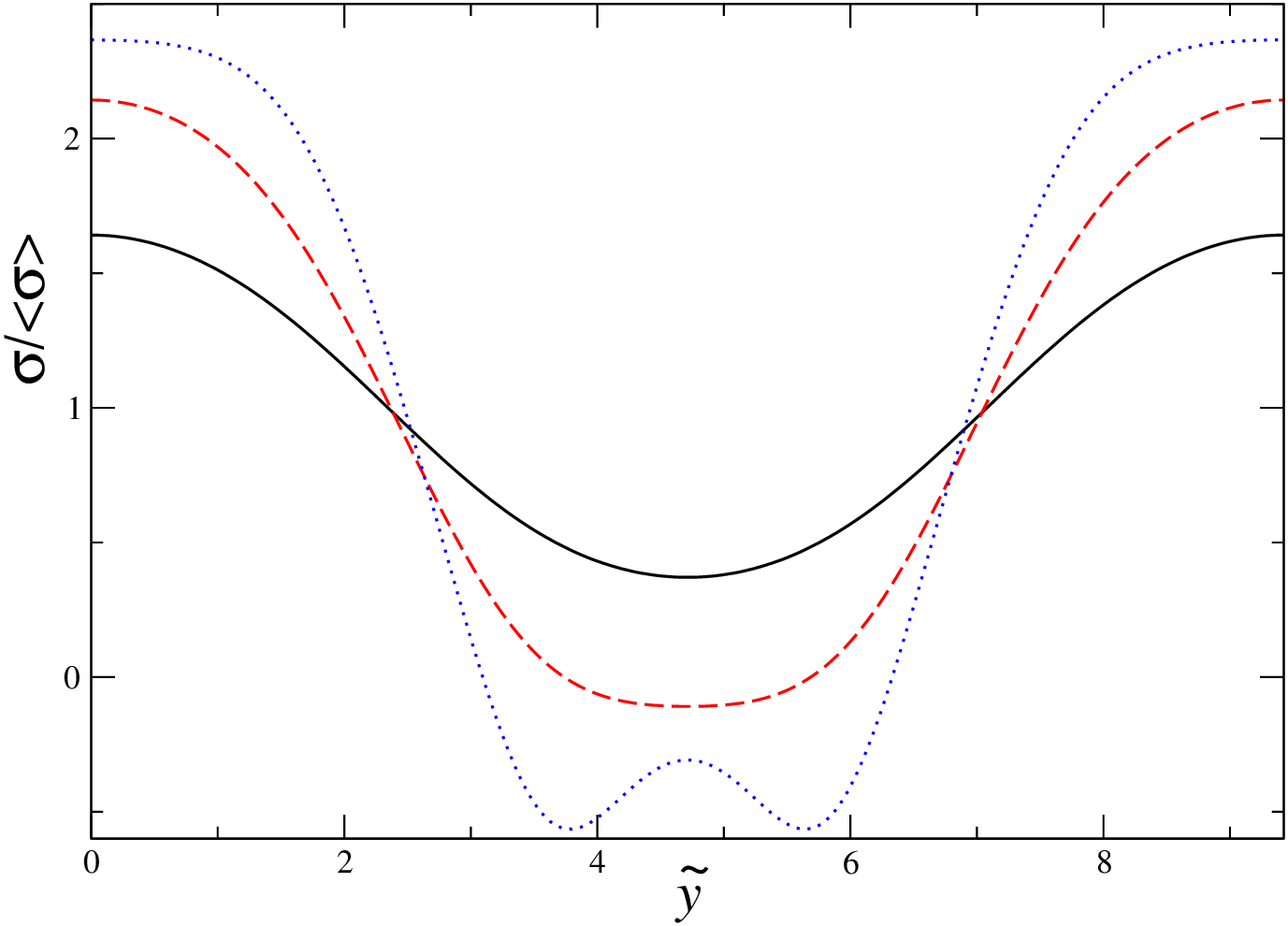}
\caption{
Plots of $\sigma/\langle\sigma\rangle$ on $\widetilde{y}$
for the distance between the walls $\widetilde{d}=\pi/8$
and the period parameter $b=2/3$, plotted within one period $2\pi/b=3\pi$.  
The solid line corresponds to the amplitude parameter $a=1/5$,
the dashed line to $a=2/5$ and the dotted line to $a=3/5$.}
\label{fig7}
\end{center}
\end{figure}

For $\widetilde{d}=\pi/8$ and $b=2/3$, the normalized modulated
surface charge density $\sigma/\langle\sigma\rangle$ as a function
of $\widetilde{y}$ is always positive for $a=1/5$ (solid curve)
and negative in an interval of $\widetilde{y}$ for $a=2/5$ (dashed curve)
and $a=3/5$ (dotted curve), see Fig. \ref{fig7}.

Explicit analytic results show that for our family of
exactly solvable models in the PB limit, surface charge modulation can
both increase (small distances) and decrease (large distances)
the pressure between two symmetrically charged walls.

\renewcommand{\theequation}{8.\arabic{equation}}
\setcounter{equation}{0}

\section{Conclusion} \label{conclusion}
The exact results are of interest in statistical mechanics because they
unambiguously answer important questions without any hypothesis.
We tried to answer the question of how the modulation of the surface charge
affects the force (or equivalently the pressure) between two symmetric walls
at distance $d$.
We restricted ourselves to the mean-field PB theory valid in the high
temperature region.
Previous approaches based on a combination of analytical perturbation methods
and MC simulations \cite{Lukatsky02a,Henle04,Lukatsky02b,Khan05}
suggest that the modulation of the surface charge density causes a decrease
in the force between the walls.
The argument is that an increase in the particle density near the walls with
modulated surface charge causes a decrease in the particle density in
the midplane between the walls; since the force (pressure) is proportional
to this particle density, it decreases.
However, as is shown in Eq. (\ref{P0}), the local pressure has an additional
positive contribution due to the variation of the potential along
the midplane, so increasing or decreasing the pressure is questionable.
To obtain an exact solution to the problem, we restrict ourselves to
modulating the surface charge density along only one direction.
This simplification leads to a 2D Liouville equation for the electric
potential (\ref{Poissonprime}) with boundary conditions (\ref{bcprime}),
for which the most general solution exists \cite{Crowdy97},
see Eqs. (\ref{0}) -- (\ref{3}).
Appropriate choice of the trial functions $Y_1(z)$, $Y_2(z)$ and the constants
$c_1$, $c_2$ and $c_3$ leads to a relatively complicated result for
the potential given by Eqs. (\ref{fii}) -- (\ref{hh}).
The surface charge density is generated inversely
from the electrostatic potential, so our exact results hold
for a limited set of models with a specific variation of
the surface charge density with the inter-wall distance $d$.
The exact potential is real and free of singularities under a series of
conditions for the dimensionless distance between walls $\widetilde{d}$,
the amplitude and period parameters $a$ and $b$ of the potential, derived in
section \ref{conditions}.
In practical calculations in the next section \ref{analysis},
these conditions restrict the allowed values of $\widetilde{d}$, $a$ and $b$
only slightly.
The results for the ratio of the dimensionless pressure $\bar{P}$ for
the model with surface charge modulation and the pressure $\bar{P}_0$
for the model with the uniform surface charge density equal to the mean value
of the modulated surface charge density $\langle\sigma\rangle$ given by
the relation (\ref{surfcharge}) are shown for $\widetilde{d}=\pi/2$
in Fig. \ref{fig2}, for $\widetilde{d}=\pi/4$ in Fig. \ref{fig4}, and for
$\widetilde{d}=\pi/8$ in Fig. \ref{fig6}.
It is clear from these figures that at larger distances, surface charge
modulation reduces the force between the walls, while at smaller distances,
modulation enhances this force.
This is the most important result of this work.
The corresponding plots of the normalized surface charge density pictured in
Figs. \ref{fig3}, \ref{fig5} and \ref{fig7} show that if the amplitude
parameter $a$ is small, the surface charge density is positive, while
in certain regions of the dimensionless coordinate $\widetilde{y}$ it can be
negative if $a$ is large.
It is important to note that for potential parameters
$a=1/5$, $b=2/3$ and dimensionless distances $\tilde{d}=\pi/4, \pi/8$,
there is an increase in pressure due to modulation of the surface charge
density even for ``physical'' charge distributions with
$\sigma(\tilde{y})\ge 0$.

From the perspective of further possibilities of exploring the present formalism
in the future, it seems interesting to look for exact solutions of
the 2D Liouville equation that lead to non-symmetric cases of surface
charge distribution on walls.
This would allow solving, for example, the problem of the impact of
a phase shift between modulated surface charges on the force between
the walls.
Another possibility is to find solutions of the 2D Liouville equation that
generate the surface charge as a sum of Dirac delta functions.

In a previous work \cite{Samaj19}, which dealt with a single wall with surface
charge density modulation, exact soliton solutions of the 2D sine-Gordon
equation, which represents the PB theory of the Coulomb plasma of $\pm$ charges,
were found.
It would be interesting to derive analogous soliton solutions for
the two-component plasma between two parallel walls.

\begin{acknowledgements}
This work was supported by the Slovak Research and Development Agency under
the Contract no. APVV-24-0091 and VEGA Grant no. 2/0089/24.
\end{acknowledgements}

\section*{Declarations}

\begin{itemize}
\item Funding   Not applicable 
\item Conflict of interest/Competing interests   Not applicable
\item Ethics approval and consent to participate   Not applicable
\item Consent for publication   Not applicable   
\item Data availability   Data are available upon a request. 
\item Materials availability   Not applicable
\item Code availability   Not applicable 
\item Author contribution   Not applicable
\end{itemize}


\begin{thebibliography}{10}

\bibitem{Raspaud00} Raspaud, E., da Conceicao, M., Livolant, F.:
Do free DNA counterions control the osmotic pressure?
Phys. Rev. Lett. {\bf 84}, 2533--2536 (2000)
  
\bibitem{Palberg04} Palberg, T., Medebach, M., Garbow, N., Evers, M.,
Fontecha, A.B., Reiber, H., Bartsch, E.: 
Electrophoresis of model colloidal spheres in low salt aqueous suspension.
J. Phys.: Condens. Matter. {\bf 16}, S4039--S4050 (2004)

\bibitem{Brunner04} Brunner, M., Dobnikar, J., von Gr\"{u}nberg, H.H.,
Bechinger, C.:
Direct measurement of three-body interactions amongst charged colloids.  
Phys. Rev. Lett. {\bf 92}, 078301 (2004)

\bibitem{Attard96} Attard, Ph.: 
Electrolytes and the electric double layer.
Adv. Chem. Phys. {\bf XCII}, 1--159 (1996)

\bibitem{Levin02} Levin, Y.: 
Electrostatic correlations: from Plasma to Biology.
Rep. Prog. Phys. {\bf 65}, 1577--1632 (2002)

\bibitem{Messina09} Messina, R.: 
Electrostatics in soft matter.
J. Phys.: Condens. Matter {\bf 21}, 113102 (2009)

\bibitem{Hansen00} Hansen, J.P., L\"owen, H.:
Effective interactions between electric double layers.
Annu. Rev. Phys. Chem. {\bf 51}, 209--242 (2000)

\bibitem{Gulbrand84} Gulbrand, L., J\"onsson, B., Wennerstr\"om, H., Linse, P.: 
Electrical double layer forces. A Monte Carlo study.
J. Chem. Phys. {\bf 80}, 2221-2228 (1984)

\bibitem{Kjellander84} Kjellander, R., Mar\v{c}elja, S.: 
Correlation and image charge effects in electric double-layers.
Chem. Phys. Lett. {\bf 112}, 49--53 (1984)

\bibitem{Gronbech97} Gr{\o}nbech-Jensen, N., Mashl, R.J., Bruinsma, R.F.,
Gelbart, W.M.: 
Counterion-Induced attraction between rigid polyelectrolytes.
Phys. Rev. Lett. {\bf 78}, 2477--2480 (1997) 

\bibitem{Khan85} Khan, A., J\"onsson, B., Wennerstr\"om, H.: 
Phase equilibria in the mixed sodium and calcium di-2-ethylhexylsulfosuccinate 
aqueous system. An illustration of repulsive and attractive double-layer forces.
J. Phys. Chem. {\bf 89}, 5180-5184 1985

\bibitem{Kjellander88} Kjellander, R., Mar\v{c}elja, S., Quirk, J.P.: 
Attractive double-layer interactions between calcium clay particles.
J. Colloid Interface Sci. {\bf 126}, 194--211 (1988)

\bibitem{Bloomfield91} Bloomfield, V.A.: 
Condensation of DNA by multivalent cations: Considerations on mechanism.
Biopolymers {\bf 31}, 1471--1481 (1991)

\bibitem{Kekicheff93} K\'ekicheff, P., Mar\v{c}elja, S., Senden, T.J., 
Shubin, V.E.: 
Charge reversal seen in electrical double layer interaction of surfaces 
immersed in 2:1 calcium electrolyte. 
J. Chem. Phys. {\bf 99}, 6098--6113 (1993)

\bibitem{Dubois98} Dubois, M., Zemb, T., Fuller, N., Rand, R.P., 
Pargesian, V.A.: 
Equation of state of a charged bilayer system: Measure of the entropy of 
the lamellar–lamellar transition in DDABr. 
J. Chem. Phys. {\bf 108}, 7855--7869 (1998)


\bibitem{Attard88} Attard, P., Mitchell, D.J., Ninham, B.W.:
Beyond Poisson-Boltzmann: Images and correlations in the electric double
layer. I. Counterions only.
J. Chem. Phys. {\bf 88}, 4987--4996 (1988)

\bibitem{Podgornik90} Podgornik, R.: 
An analytic treatment of the first-order correction to the Poisson-Boltzmann
interaction free energy in the case of counter-ion only Coulomb fluid.
J. Phys. A: Math. Gen. {\bf 23}, 275--284 (1990)

\bibitem{Netz00} Netz, R.R., Orland, H.:
Beyond Poisson-Boltzmann: Fluctuation effects and correlation functions.
Eur. Phys. J. E {\bf 1}, 203--214 (2000)

\bibitem{Moreira01} Moreira, A.G., Netz, R.R.: 
Binding of similarly charged plates with counterions only.
Phys. Rev. Lett. {\bf 87}, 078301 (2001)

\bibitem{Netz01} Netz, R.R.: 
Electrostatics of counter-ions at and between planar charged walls: from
Poisson-Boltzmann to the strong-coupling theory.
Eur. Phys. J. E {\bf 5}, 557--574 (2001)

\bibitem{Boroudjerdi05} Boroudjerdi, H., Kim, Y-W., Naji, A., Netz, R.R., 
Schlagberger, X., Serr, A.: 
Statics and dynamics of strongly charged soft matter.
Phys. Rep. {\bf 416}, 129--199 (2005)

\bibitem{Kanduc07} Kandu\v{c}, M., Podgornik, R.: 
Electrostatic image effects for counterions between charged planar walls.
Eur. Phys. J. E {\bf 23}, 265--274 (2007)

\bibitem{Samaj11} \v{S}amaj, L., Trizac, E.: 
Counterions at highly charged interfaces: From one plate to like-charge
attraction.
Phys. Rev. Lett. {\bf 106}, 078301 (2011)

\bibitem{Samaj16} \v{S}amaj, L., dos Santos, A.P., Levin, Y., Trizac, E.:
Mean-field beyond mean-field: the single particle view for moderately 
to strongly coupled charged fluids.
Soft Matter {\bf 12}, 8768--8773 (2016)

\bibitem{Chan80} Chan, D.Y.C, Mitchell, J., Ninham, B.W.: 
A self‐consistent study of ion adsorption and discrete charge effects in
the electrical double layer.
J. Chem. Phys. {\bf 72}, 5159--5162 (1980)

\bibitem{Gonzalez01} Gonzalez-Amezcua, O., Hernandez-Contreras, M.,
Pincus, P.A.: 
Electrostatic correlation force of discretely charged membranes.  
Phys. Rev. E {\bf 64}, 041603 (2001)

\bibitem{Lukatsky02a} Lukatsky, D.B., Safran, S.A, Lau, A.W.C, Pincus, P.A.: 
Enhanced counterion localization induced by surface charge modulation.
Europhys. Lett. {\bf 58}, 785--791 (2002)

\bibitem{Henle04} Henle, M.L, Santangelo, C.D., Patel, D.M., Pincus, P.A.: 
Distribution of counterions near discretely charged planes and rods.
Europhys. Lett. {\bf 66}, 284--290 (2004) 

\bibitem{Lukatsky02b} Lukatsky, D.B., Safran, S.A.: 
Universal reduction of pressure between charged surfaces by long-wavelength
surface charge modulation.
Europhys. Lett. {\bf 60}, 629--635 (2002)

\bibitem{Khan05} Khan, M.O., Petris, S., Chan, D.Y.C.: 
The influence of discrete surface charges on the force between charged
surfaces.
J. Chem. Phys. {\bf 122}, 104705 (2005)

\bibitem{Samaj22} \v{S}amaj, L.:
Electric double layers with modulated surface charge density: exact 2D
results.
J. Phys. A: Math. Theor. {\bf 55}, 275001 (2022)

\bibitem{Henderson78} Henderson, D., Blum, L.: 
Some exact results and the application of the mean spherical approximation 
to charged hard spheres near a charged hard wall.
J. Chem. Phys. {\bf 69}, 5441--5449 (1978)

\bibitem{Henderson79} Henderson, D., Blum, L., Lebowitz, J.L., 
An exact formula for the contact value of the density profile of 
a system of charged hard spheres near a charged wall.
J. Electroanal. Chem. {\bf 102}, 315--319 (1979)

\bibitem{Blum81} Blum, L., Henderson, D., Lebowitz, J.L., Gruber, Ch., 
Martin, Ph.A.: 
A sum rule for an inhomogeneous electrolyte.
J. Chem. Phys. {\bf 75}, 5974--5975 (1981) 

\bibitem{Wennerstrom82} Wennerstr\"om, H., J\"onsson, B., Linse, P.: 
The cell model for polyelectrolyte systems. Exact statistical mechanical
relations, Monte Carlo simulations, and the Poisson-Boltzmann approximation.  
J. Chem. Phys. {\bf 76}, 4665--4670 (1982)

\bibitem{Samaj25} \v{S}amaj, L.:
Contact value theorem for electric double layers with modulated surface
charge density.  
J. Phys. A: Math. Theor. {\bf 58}, 065001 (2025)

\bibitem{Gouy10} Gouy, G.L.: 
Sur la constitution de la charge \'{e}lectrique \`{a} la surface d'un
\'{e}lectrolyte.
J. Phys. Theor. Appl. {\bf 9}, 457--468 (1910)

\bibitem{Chapman13} Chapman, D.L.: 
A contribution to the theory of electrocapillarity.
Philos. Mag. {\bf 25}, 475--481 (1913)

\bibitem{Samaj19} \v{S}amaj, L., Trizac, E.: 
Electric double layers with surface charge modulations:
Exact Poisson-Boltzmann solutions.
Phys. Rev. E {\bf 100}, 042611 (2019)

\bibitem{Crowdy97} Crowdy, D.G.: 
General solutions to the 2D Liouville equation.  
Int. J. Eng. Sci. {\bf 35}, 141--149 (1997)

\bibitem{Jackson98} Jackson, J.D.: 
Classical Electrodynamics, 3rd edn.,
John Wiley, (1998)

\bibitem{Andelman06} Andelman, D.: 
Introduction to electrostatics in soft and biological matter
In: Poon, W.C.K., Andelman, D. (eds.), Soft Condensed Matter Physics
in Molecular and Cell Biology, vol. 6, Taylor \& Francis, (2006)

\end{thebibliography}
\end{document}